%% file: main.tex
\documentclass[conference,compsoc]{IEEEtran}
\ifCLASSOPTIONcompsoc
  \usepackage[nocompress]{cite}
\else
  \usepackage{cite}
\fi
\ifCLASSINFOpdf
\else
\fi
\usepackage{xurl}
\usepackage{hyperref}
\usepackage{booktabs}
\usepackage{xspace}
\usepackage{enumitem}
\usepackage{makecell}
\usepackage{xcolor}
\usepackage{amssymb}
\usepackage{amsmath}
\usepackage{graphicx}
\usepackage{placeins} 

\usepackage{xparse}

\hypersetup{
  colorlinks,
  linkcolor={red!80!black},
  citecolor={blue},
  urlcolor={blue!80!black}
}

\newcommand{\mcpsec}{\textsc{MCPSec}\xspace}
\newcommand{\IPI}{IPI\xspace}

\newcommand{\ipi}{indirect prompt injection\xspace}
\newcommand{\mcp}{\textsc{MCP}\xspace}
\newcommand{\NB}{No-box\xspace}
\newcommand{\nb}{no-box\xspace}
\newcommand{\ignore}[1]{}

\newcommand{\numMCPs}{20}
\newcommand{\costPerServer}{\$1.65}

\newcommand{\medianStars}{3,764}
\newcommand{\meanStars}{9,479}

\newcommand{\numSampledTools}{177}
\newcommand{\numFlows}{143}
\newcommand{\numVulTools}{95}

\newcommand{\stageOnePlausiblePct}{96.5\%} 
\newcommand{\stageOneDataFidelityPct}{83.9\%} 
\newcommand{\stageOneSemanticExecPct}{86.7\%} 
\newcommand{\stageOneAttackPreqPct}{86.7\%} 
\newcommand{\stageOneSanitizationPct}{76.9\%} 
\newcommand{\stageOneTocCoherencePct}{85.3\%} 
\newcommand{\stageOneMeanAgreementPct}{83.6\%} 

\newcommand{\numBaselineMatchedFlows}{97}

\newcommand{\numAgentBaselineCandidates}{98}
\newcommand{\numAgentBaselineMatchedFlows}{97}
\newcommand{\agentBaselineDataFidelityPct}{62.9\%} 
\newcommand{\agentBaselineSemanticExecPct}{90.7\%} 
\newcommand{\agentBaselineAttackPreqPct}{89.7\%} 
\newcommand{\agentBaselineSanitizationPct}{66.0\%} 
\newcommand{\agentBaselineMeanAgreementPct}{77.3\%} 
\newcommand{\stageOneAgentMatchedDataFidelityPct}{82.5\%} 
\newcommand{\stageOneAgentMatchedSemanticExecPct}{92.8\%} 
\newcommand{\stageOneAgentMatchedAttackPreqPct}{93.8\%} 
\newcommand{\stageOneAgentMatchedSanitizationPct}{77.3\%} 
\newcommand{\stageOneAgentMatchedMeanAgreementPct}{86.6\%} 

\newcommand{\sanitizationDisagreementNum}{31} 
\newcommand{\downgradedAttackPrereq}{11} 

\newcommand{\FeasibleReachableNum}{96}
\newcommand{\FeasibleReachablePct}{91.4\%}
\newcommand{\NotFeasibleButReachableNum}{25}
\newcommand{\NotFeasibleButReachablePct}{71.4\%}
\newcommand{\NotFeasibleReachableLowFidelityPct}{88.0\%} 
\newcommand{\FeasibleReachableLowFidelityPct}{27.1\%} 
\newcommand{\NotFeasibleReachableLowExecPct}{68.0\%} 
\newcommand{\FeasibleReachableLowExecPct}{14.6\%} 

\newcommand{\stageTwoDataFidelityPct}{67.1\%}    
\newcommand{\stageTwoSemanticExecPct}{62.9\%}    
\newcommand{\stageTwoSanitizationPct}{69.2\%}    
\newcommand{\stageTwoMeanAgreementPct}{66.4\%}   
\newcommand{\agentBaselineStageTwoDataFidelityPct}{60.8\%} 
\newcommand{\agentBaselineStageTwoSemanticExecPct}{63.9\%} 
\newcommand{\agentBaselineStageTwoSanitizationPct}{63.9\%} 
\newcommand{\agentBaselineStageTwoMeanAgreementPct}{62.9\%} 
\newcommand{\stageTwoAgentMatchedDataFidelityPct}{70.1\%} 
\newcommand{\stageTwoAgentMatchedSemanticExecPct}{67.0\%} 
\newcommand{\stageTwoAgentMatchedSanitizationPct}{72.2\%} 
\newcommand{\stageTwoAgentMatchedMeanAgreementPct}{69.8\%} 

\newcommand{\numSharedBaselineFlows}{95}
\newcommand{\sharedMcpsecDataFidelityPct}{84.2\%} 
\newcommand{\sharedMcpsecSemanticExecPct}{93.7\%} 
\newcommand{\sharedMcpsecAttackPreqPct}{93.7\%} 
\newcommand{\sharedMcpsecSanitizationPct}{78.9\%} 
\newcommand{\sharedMcpsecMeanAgreementPct}{87.6\%} 
\newcommand{\sharedBaselineDataFidelityPct}{58.9\%} 
\newcommand{\sharedBaselineSemanticExecPct}{89.5\%} 
\newcommand{\sharedBaselineAttackPreqPct}{85.3\%} 
\newcommand{\sharedBaselineSanitizationPct}{63.2\%} 
\newcommand{\sharedBaselineMeanAgreementPct}{74.2\%} 
\newcommand{\sharedAgentDataFidelityPct}{64.2\%} 
\newcommand{\sharedAgentSemanticExecPct}{90.5\%} 
\newcommand{\sharedAgentAttackPreqPct}{91.6\%} 
\newcommand{\sharedAgentSanitizationPct}{67.4\%} 
\newcommand{\sharedAgentMeanAgreementPct}{78.4\%} 
\newcommand{\sharedMcpsecStageTwoDataFidelityPct}{69.5\%} 
\newcommand{\sharedMcpsecStageTwoSemanticExecPct}{68.4\%} 
\newcommand{\sharedMcpsecStageTwoSanitizationPct}{73.7\%} 
\newcommand{\sharedMcpsecStageTwoMeanAgreementPct}{70.5\%} 
\newcommand{\sharedBaselineStageTwoDataFidelityPct}{52.6\%} 
\newcommand{\sharedBaselineStageTwoSemanticExecPct}{62.1\%} 
\newcommand{\sharedBaselineStageTwoSanitizationPct}{62.1\%} 
\newcommand{\sharedBaselineStageTwoMeanAgreementPct}{58.9\%} 
\newcommand{\sharedAgentStageTwoDataFidelityPct}{60.0\%} 
\newcommand{\sharedAgentStageTwoSemanticExecPct}{65.3\%} 
\newcommand{\sharedAgentStageTwoSanitizationPct}{65.3\%} 
\newcommand{\sharedAgentStageTwoMeanAgreementPct}{63.5\%} 
\newcommand{\PlausibleSourcePct}{93.7\%}
\newcommand{\PlausibleSourceNum}{134}
\newcommand{\ImplausibleSourceNum}{9}
\newcommand{\NoSanitizationPct}{92.3\%}
\newcommand{\NoSanitizationNum}{132}
\newcommand{\PartialSanitizationPct}{3.5\%}
\newcommand{\PartialSanitizationNum}{5}
\newcommand{\EffectiveSanitizationPct}{4.2\%}
\newcommand{\EffectiveSanitizationNum}{6}
\newcommand{\FeasibleToCPct}{73.4\%}
\newcommand{\FeasibleToCNum}{105}
\newcommand{\PartialFeasibleToCPct}{2.1\%}
\newcommand{\PartialFeasibleToCNum}{3}
\newcommand{\UnfeasibleToCPct}{24.5\%}
\newcommand{\UnfeasibleToCNum}{35}
\newcommand{\combinedFeasiblePartialFeasibleToCPct}{75.5\%}
\newcommand{\combinedUnfeasiblePartialFeasibleToCNum}{38}

\newcommand{\ImplausibleSourceUnfeasibleToCNum}{9}

\newcommand{\FactorWrongSourceNum}{9}
\newcommand{\FactorWrongSourcePct}{23.7\%}
\newcommand{\FactorDataNotPreservedNum}{24}
\newcommand{\FactorDataNotPreservedPct}{63.2\%}
\newcommand{\FactorEffectiveSanitizationNum}{6}
\newcommand{\FactorEffectiveSanitizationPct}{15.8\%}
\newcommand{\FactorOutputTransformedNum}{22}
\newcommand{\FactorOutputTransformedPct}{57.9\%}

\newcommand{\ReachableVerbatimPct}{11.2\%}
\newcommand{\ReachableVerbatimNum}{16}
\newcommand{\ReachableModifiedPct}{61.5\%}
\newcommand{\ReachableModifiedNum}{88}
\newcommand{\ReachableMetadataPct}{14.0\%}
\newcommand{\ReachableMetadataNum}{20}

\newcommand{\NotReachablePct}{13.3\%}
\newcommand{\NotReachableNum}{19}
\newcommand{\TotalInstructionReachablePct}{72.7\%} 
\newcommand{\TotalInstructionReachableNum}{104} 

\newcommand{\humanAgreementOverallPct}{83.3\%} 

\newcommand{\RQfourUnlabeled}{82}          

\newcommand{\RQfourMcpsecTP}{94}

\newcommand{\RQfourMcpsecFN}{1}

\newcommand{\RQfourMcpsecRecall}{98.9\%}

\newcommand{\RQfourBaseTP}{80}

\newcommand{\RQfourBaseRecall}{84.2\%}

\newcommand{\RQfourAgentTP}{81}

\newcommand{\RQfourAgentFN}{14}

\newcommand{\RQfourAgentRecall}{85.3\%}

\newcommand{\RQfourAgentFNCollab}{12}

\input{sections/tables.tex}

\begin{document}
%
\title{No-Box Vulnerability Analysis: Description-Only Detection of Indirect Prompt Injection Vulnerabilities in MCP Servers}


\author{
\IEEEauthorblockN{
Zehua Zhang\textsuperscript{*}, Jie Hu, Pratham Hegde, Aditya Maheshbhai Gabani,
Souradip Nath, Yibo Liu, Siyu Liu, \\ Hongkai Chen,
Hulin Wang, Zhuoer Lyu, Chang Zhu, Divij Handa,
Yan Shoshitaishvili, Tiffany Bao, \\ Ruoyu Wang, Adam Doupé
}
\IEEEauthorblockA{
School of Computing and Augmented Intelligence, Arizona State University \\
\textsuperscript{*}Correspondence: \href{mailto:zzhan645@asu.edu}{zzhan645@asu.edu}
}
}


%


\maketitle

\input{sections/00_abstract}


%
\IEEEpeerreviewmaketitle

\input{sections/01_introduction}
\input{sections/02_background}

\input{sections/03_nobox}
\input{sections/04_caseMCP}

\input{sections/05_system_design}

\input{sections/06_experiment}
\input{sections/07_evaluation}
\input{sections/08_discussion}

\input{sections/09_related_work}
\input{sections/10_conclusion}

\newpage
\bibliographystyle{IEEEtran}
%
\bibliography{references}

\clearpage
\appendices
\input{sections/appendix/ethics}                  
\input{sections/appendix/model_sensitivity}       
\input{sections/appendix/baselines}               
\input{sections/appendix/vulnerability_labeling}  
\end{document}

%% file: sections/tables.tex
\newcommand{\RiskRubricTable}{
\begin{table}[tb]
\centering
\small
\setlength{\tabcolsep}{3pt}
\caption{Four-axis risk assessment rubric.}
\label{tab:risk-rubric}
\begin{tabular}{@{}ll@{}}
\toprule
\textbf{Axis / Level} & \textbf{Criterion or Examples} \\
\midrule
\multicolumn{2}{@{}l}{\textit{Payload Fidelity}} \\
\quad \textsc{high} (verbatim) & Source content preserved intact \\
\quad \textsc{low} (derived) & Content summarized or reduced \\[4pt]
\multicolumn{2}{@{}l}{\textit{Attacker Controllability}} \\
\quad \textsc{high} (public/supply-chain) & No authentication needed \\
\quad \textsc{med} (authenticated remote) & Platform account or role required \\
\quad \textsc{low} (implausible) & Requires local access \\[4pt]
\multicolumn{2}{@{}l}{\textit{Semantic Expressiveness}} \\
\quad \textsc{high} (expressive) & Free-form text \\
\quad \textsc{low} (constrained) & Hashes, IDs, or enums \\[4pt]
\multicolumn{2}{@{}l}{\textit{Sanitization}} \\
\quad \textsc{high} (none) & No explicit defense \\
\quad \textsc{med} (partial) & Truncation or length limits \\
\quad \textsc{low} (effective) & Escaping, filtering, or redaction \\
\bottomrule
\end{tabular}
\end{table}
}
\newcommand{\EvalTargetTable}{
\begin{table*}[tb]
  \centering
  \caption{Evaluation target MCP servers, grouped by category and sorted by tool count (descending) within each
  category. GitHub stars as of April~2026. \emph{LoC}: source lines of code counted with \texttt{cloc}, excluding tests, fixtures, documentation, skills, and dependencies.
  \emph{Paid}: whether the upstream API requires a paid plan to test
  all sampled tools.}
  \resizebox{\textwidth}{!}{
  \begin{tabular}{@{}rlllrrrll@{}}
  \toprule
  \textbf{\#} & \textbf{Server} & \textbf{Repository} & \textbf{Commit} & \textbf{Tools} & \textbf{LoC} & \textbf{Stars}
  & \textbf{Provider} & \textbf{Paid Access} \\
  \midrule
    \multicolumn{9}{@{}l}{\textbf{A: Search, Web, and Document Retrieval} (47 total tools) --- fetch public web content,
  search results, or documents} \\[2pt]
    1 & Firecrawl & \texttt{firecrawl/firecrawl-mcp-server} & \texttt{6fef044} & 14 & 1,152 & 6,116 & Official & No \\
    2 & arXiv & \texttt{blazickjp/arxiv-mcp-server} & \texttt{51ebf3a} & 10 & 1,966 & 2,573 & Community & No \\
    3 & Exa Search & \texttt{exa-labs/exa-mcp-server} & \texttt{8df65b0} & 10 & 2,381 & 4,286 & Official & No \\
    4 & Brave Search & \texttt{brave/brave-search-mcp-server} & \texttt{52590e9} & 6 & 2,586 & 924 & Official & Yes \\
    5 & Tavily & \texttt{tavily-ai/tavily-mcp} & \texttt{238f6fd} & 5 & 791 & 1,812 & Official & Yes \\
    6 & Context7 & \texttt{upstash/context7} & \texttt{658ec67} & 2 & 715 & 53,372 & Official & No \\
  \midrule
    \multicolumn{9}{@{}l}{\textbf{B: Browser-Mediated Web Interaction} (117 total tools) --- observe or manipulate live
  browser state} \\[2pt]
    7 & Skyvern & \texttt{Skyvern-AI/skyvern} & \texttt{e565094} & 49 & 6,207 & 21,314 & Official & No \\
    8 & Playwright & \texttt{executeautomation/mcp-playwright} & \texttt{2349c28} & 33 & 3,929 & 5,455 & Community & No
  \\
    9 & Chrome DevTools & \texttt{ChromeDevTools/chrome-devtools-mcp} & \texttt{b1684c6} & 29 & 8,799 & 36,597 &
  Official & No \\
    10 & Browserbase & \texttt{browserbase/mcp-server-browserbase} & \texttt{f6bd321} & 6 & 1,356 & 3,277 & Official &
  Yes \\
  \midrule
    \multicolumn{9}{@{}l}{\textbf{C: SaaS Collaboration and Communication} (388 total tools) --- read user-authored
  content via authenticated APIs} \\[2pt]
    11 & GitLab & \texttt{zereight/gitlab-mcp} & \texttt{c393d1e} & 117 & 10,053 & 1,392 & Community & No \\
    12 & Google Workspace & \texttt{taylorwilsdon/google\_workspace\_mcp} & \texttt{35fcc94} & 114 & 18,967 & 2,180 &
  Community & Yes \\
    13 & Atlassian & \texttt{sooperset/mcp-atlassian} & \texttt{8e84d74} & 73 & 20,621 & 4,993 & Community & No \\
    14 & GitHub & \texttt{github/github-mcp-server} & \texttt{2a1eaac} & 41 & 19,437 & 29,151 & Official & No \\
    15 & Notion & \texttt{makenotion/notion-mcp-server} & \texttt{3bef7ad} & 22 & 1,170 & 4,250 & Official & No \\
    16 & Slack & \texttt{korotovsky/slack-mcp-server} & \texttt{b24b1b0} & 12 & 7,119 & 1,550 & Community & No \\
    17 & Discord (Klavis) & \texttt{Klavis-AI/Klavis} & \texttt{340b6cf} & 9 & 639 & 5,715 & Community & No \\
  \midrule
    \multicolumn{9}{@{}l}{\textbf{D: Infrastructure, Cloud, and Data Platforms} (60 total tools) --- read operational,
  registry, or project-state data} \\[2pt]
    18 & Supabase & \texttt{supabase-community/supabase-mcp} & \texttt{1cd04f0} & 29 & 14,581 & 2,633 & Community & Yes
  \\
    19 & Sentry & \texttt{getsentry/sentry-mcp} & \texttt{0941c50} & 22 & 16,093 & 658 & Official & No \\
    20 & Terraform & \texttt{hashicorp/terraform-mcp-server} & \texttt{617ba91} & 9 & 7,200 & 1,334 & Official & No \\
  \bottomrule
  \end{tabular}
  }
  \label{tab:eval-targets}
\end{table*}
}

\newcommand{\EvalOutlineTable}{%
\begin{table*}[tb]
\centering
\small
\caption{Evaluation progression from metadata-only assessment to controlled vulnerability validation.}
\label{tab:eval-progression}
\begin{tabular}{@{}llll@{}}
\toprule
\textbf{Stage}  & \textbf{Evidence} & \textbf{Annotation} &
\textbf{Purpose} \\
\midrule
RQ1  &
Metadata &
ToC plausibility and risk rubric agreement &
Human \nb{} assessment \\

RQ2  &
Metadata + source code &
ToC plausibility and risk rubric correctness &
Static analysis validation \\

RQ3 &
Metadata + source code + runtime &
Reachability and runtime fidelity &
Dynamic data flow reachability test \\

RQ4  &
PoCs for suitable tools &
Tool vulnerability labels &
Vulnerability analysis performance \\
\bottomrule
\end{tabular}
\end{table*}
}

\newcommand{\RQOneResultsTable}{%
\begin{table}[tb]
\centering
\caption{RQ1 results using registration metadata alone. We report \mcpsec's performance on the \numFlows{} identified flows and compare it with that of the agentic baseline on the \numAgentBaselineMatchedFlows{} overlapping tools flagged by both methods.}
\resizebox{\columnwidth}{!}{%
\begin{tabular}{@{}lccc@{}}
\toprule
\textbf{Measure}
  & \shortstack{\textbf{\mcpsec}\\($N=\numFlows$)}
  & \shortstack{\textbf{\mcpsec}\\($N=\numAgentBaselineMatchedFlows$)}
  & \shortstack{\textbf{Agentic}\\($N=\numAgentBaselineMatchedFlows$)} \\
\midrule
\multicolumn{4}{@{}l}{\textit{Metadata-grounded plausibility}} \\
External source plausible
  & \stageOnePlausiblePct & -- & -- \\
ToC plausible
  & \stageOneTocCoherencePct & -- & -- \\
\addlinespace[2pt]
\multicolumn{4}{@{}l}{\textit{Agreement with human risk labels}} \\
\texttt{Payload\_Fidelity}
  & \stageOneDataFidelityPct & \stageOneAgentMatchedDataFidelityPct & \agentBaselineDataFidelityPct \\
\texttt{Semantic\_Expressiveness}
  & \stageOneSemanticExecPct & \stageOneAgentMatchedSemanticExecPct & \agentBaselineSemanticExecPct \\
\texttt{Sanitization}
  & \stageOneSanitizationPct & \stageOneAgentMatchedSanitizationPct & \agentBaselineSanitizationPct \\
\texttt{Attacker\_Controllability}
  & \stageOneAttackPreqPct & \stageOneAgentMatchedAttackPreqPct & \agentBaselineAttackPreqPct \\
\textbf{Mean (four axes)}
  & \textbf{\stageOneMeanAgreementPct} & \textbf{\stageOneAgentMatchedMeanAgreementPct} & \textbf{\agentBaselineMeanAgreementPct} \\
\bottomrule
\end{tabular}%
}
\label{tab:rq1-results}
\end{table}
}

\newcommand{\RQTwoResultsTable}{%
\begin{table}[tb]
\centering
\caption{RQ2 results under source-code evidence. }
\resizebox{\columnwidth}{!}{%
\begin{tabular}{@{}lccc@{}}
\toprule
\textbf{Measure}
  & \shortstack{\textbf{\mcpsec}\\($N=\numFlows$)}
  & \shortstack{\textbf{\mcpsec}\\($N=\numAgentBaselineMatchedFlows$)}
  & \shortstack{\textbf{Agentic}\\($N=\numAgentBaselineMatchedFlows$)} \\
\midrule
\multicolumn{4}{@{}l}{\textit{Source-code-grounded plausibility}} \\
External source plausible
  & \PlausibleSourceNum{} (\PlausibleSourcePct{}) & -- & -- \\
ToC: Plausible
  & \FeasibleToCNum{} (\FeasibleToCPct{}) & -- & -- \\
ToC: Partially Plausible
  & \PartialFeasibleToCNum{} (\PartialFeasibleToCPct{}) & -- & -- \\
ToC: Implausible
  & \UnfeasibleToCNum{} (\UnfeasibleToCPct{}) & -- & -- \\
\addlinespace[2pt]
\multicolumn{4}{@{}l}{\textit{Agreement with human risk labels}} \\
\texttt{Payload\_Fidelity}
  & \stageTwoDataFidelityPct & \stageTwoAgentMatchedDataFidelityPct & \agentBaselineStageTwoDataFidelityPct \\
\texttt{Semantic\_Expressiveness}
  & \stageTwoSemanticExecPct & \stageTwoAgentMatchedSemanticExecPct & \agentBaselineStageTwoSemanticExecPct \\
\texttt{Sanitization}
  & \stageTwoSanitizationPct & \stageTwoAgentMatchedSanitizationPct & \agentBaselineStageTwoSanitizationPct \\
\textbf{Mean (three axes)}
  & \textbf{\stageTwoMeanAgreementPct} & \textbf{\stageTwoAgentMatchedMeanAgreementPct} & \textbf{\agentBaselineStageTwoMeanAgreementPct} \\
\bottomrule
\end{tabular}%
}
\label{tab:rq2-results}
\end{table}
}

\newcommand{\RQThreeResultsTable}{%
\begin{table}[tb]
\centering
\caption{RQ3 results from dynamic execution evidence.}
\begin{tabular}{@{}lrr@{}}
\toprule
\textbf{Measure} & \textbf{$n/N$} & \textbf{Percentage} \\
\midrule
\multicolumn{3}{@{}l}{\textit{Runtime verdict}} \\
Reachable-Verbatim
  & \ReachableVerbatimNum{}/\numFlows{} & \ReachableVerbatimPct \\
Reachable-Modified
  & \ReachableModifiedNum{}/\numFlows{} & \ReachableModifiedPct \\
Reachable-Metadata
  & \ReachableMetadataNum{}/\numFlows{} & \ReachableMetadataPct \\
Not Reachable
  & \NotReachableNum{}/\numFlows{} & \NotReachablePct \\
\midrule
\multicolumn{3}{@{}l}{\textit{Consistency with RQ2 ToC verdicts}} \\
RQ2 Plausible $\wedge$ RQ3 Reachable
  & \FeasibleReachableNum{}/\FeasibleToCNum{} & \FeasibleReachablePct \\
RQ2 Implausible $\wedge$ RQ3 Reachable
  & \NotFeasibleButReachableNum{}/\UnfeasibleToCNum{} & \NotFeasibleButReachablePct \\
\bottomrule
\end{tabular}
\label{tab:rq3-results}
\end{table}
}

\newcommand{\RQFourSelfContainedTable}{%
\begin{table}[tb]
\caption{RQ4 vulnerability recovery performance on the \numVulTools{} tools with confirmed \ipi{} vulnerabilities.}
\resizebox{\columnwidth}{!}{%
\begin{tabular}{@{}llrrr@{}}
\toprule
\textbf{Method} & \textbf{Base model} & \textbf{Recovered} & \textbf{Missed} & \textbf{Recall} \\
\midrule
\mcpsec & GPT-5.4
  & \RQfourMcpsecTP & \RQfourMcpsecFN & \RQfourMcpsecRecall \\
Agent Baseline & GPT-5.4
  & \RQfourAgentTP & \RQfourAgentFN & \RQfourAgentRecall \\
\bottomrule
\end{tabular}
}
\label{tab:rq4-results}
\end{table}
}

%% file: sections/00_abstract.tex
\begin{abstract}
When auditing remotely hosted services or commercially gated software for vulnerabilities, third-party analysts may lack both direct system access and the ability to interact with the running system.
We call this a \emph{\nb{} setting}, in which analysts have access only to functionality metadata that may describe the intended inputs, behaviors, and outputs of the system.
Such descriptions constrain the possible implementations of a system, enabling analysts to hypothesize potential implementations and their underlying vulnerabilities.
These hypotheses can subsequently guide validation when system access becomes available.
To demonstrate the feasibility of conducting vulnerability analysis under \nb setting, we implemented \mcpsec, a prototype that audits Model Context Protocol (MCP) servers for \ipi vulnerabilities using only the tool metadata exposed at server registration time.
Flagging vulnerable tools, \mcpsec produces hypothesized vulnerabilities and attack scenarios for later validation. 
We evaluated \mcpsec on \numMCPs{} widely deployed MCP servers comprising \numSampledTools{} tools, among which human evaluators confirmed \numVulTools{} as vulnerable.
Using only tool metadata, \mcpsec successfully identified \RQfourMcpsecTP{} of these confirmed vulnerable tools (\RQfourMcpsecRecall{} recall), compared to \RQfourAgentTP{} (\RQfourAgentRecall{} recall) identified by an agentic baseline. 
Overall, we operationalize and evaluate \nb vulnerability analysis and demonstrate its practical feasibility in realistic systems.

\end{abstract}

%% file: sections/01_introduction.tex

\section{Introduction}
\label{sec:introduction}

You're red-teaming a network, and after scanning, you identify a juicy target server.
Before you start analyzing the server, you check with your contact: uh-oh, this server is the central (and only) credit card processing server for the entire company.
Disrupting the server can result in excessive financial loss for the company, and thus you're forbidden from sending traffic to it.
What do you do?
Exclude the system from your audit? 

The issue is that traditional vulnerability discovery generally follows three paradigms: white-box analysis, where the source code of the target system is available; black-box analysis, which requires extensive interaction with the target system; and gray-box analysis, which combines interactions with information from the system.

However, there are access-constrained situations in which the source code, the runtime state, and even the ability to freely interact with the system are inaccessible or unauthorized, leaving only system metadata available. 
We refer to such situations as the \emph{\nb} setting.
The \nb setting is particularly applicable to target systems that have been developed and are in deployment, where the primary users are often third-party security analysts instead of the developers.
However, testing a deployed system for vulnerabilities can be challenging or even \emph{unethical}, as potential vulnerability discovery attempts might accidentally exploit other users.
Consider analyzing a hospital's MRI machine: no digital twin is available, a second machine cannot be procured due to expense, and the machine cannot be tested in situ or taken offline due to the impact on patient care.

To address this challenge, we propose \emph{\nb vulnerability analysis}, which only requires access to \emph{system metadata} to reason about the potential existence of vulnerabilities in a system.
Our insight is that for some vulnerability classes, even limited metadata (e.g., documentation or a system description) can suffice for hypothesizing (and eventually discovering) instances of these vulnerabilities.

Consider HotCRP, whose documentation describes its functionality that accepts an input string and queries the database for submitted papers that match.
Every implementation that is consistent with this documentation must contain a data flow from user input to a database query.

The fact that all conforming implementations have this data flow suggests a vulnerability hypothesis: the \emph{potential} existence of an SQL injection vulnerability.
We call this vulnerability hypothesis a \emph{Theory-of-Concept} (ToC), which is an explicit exploitation scenario that includes additional assumptions about the attacker's capabilities and roles and the details of the potentially vulnerable data flow.

\NB vulnerability analysis resembles security requirements engineering and threat modeling \cite{mcdermott_using_1999, van_lamsweerde_elaborating_2004, myagmar_threat_nodate, sindre_eliciting_2005}, which likewise hypothesize attacks from a system’s description instead of its implementation. 
These approaches primarily aim to identify threats and refine security requirements to guide secure system development.
Our work is conceptually inspired by them and seeks to reconstruct the security-relevant behavior of an already implemented system from only its metadata, with the goal of hypothesizing and facilitating discovery of vulnerabilities.

Of course, verifying the existence of a vulnerability requires testing it against the actual system, and a ToC can provide concrete vulnerability assumptions. A human analyst can look for evidence to support (or refute) them and correspondingly construct a Proof-of-Concept (PoC) of the vulnerability.
Alternatively, human developers (who have access to implementation details) can use the ToC to reason about potential vulnerabilities and mitigation in a timely manner.

We demonstrate the feasibility of \nb vulnerability analysis through a concrete vulnerability class: \ipi (IPI).
\IPI attacks occur when attacker-controlled instructions embedded in external content enter an LLM's context and redirect its reasoning or actions~\cite{greshake2023indirect}.
Model Context Protocol (MCP) servers provide standardized interfaces through which LLM agents access external tools and data~\cite{anthropic2024mcp}.
These interactions create opportunities for IPI when retrieved content, such as public web pages or third-party API responses, is attacker-controllable.
However, auditing MCP servers presents challenges for both white-box analysis and black-box testing. Many servers are proprietary, which limits access to their implementations. 
Live testing may require placing injection payloads in external systems, where benign users (and their LLM agents) could encounter them.
These constraints make MCP servers a suitable target for \nb analysis, which assesses potential vulnerabilities without accessing the implementation or testing exploits live.

%


We conduct the \nb vulnerability analysis
through \mcpsec, a prototype for detecting \IPI vulnerabilities in MCP servers.
Using only publicly available tool metadata, \mcpsec infers plausible data-flow paths from attacker-controllable sources to LLM input, assesses attacker capabilities against vulnerability preconditions, and produces ToCs.
Tool metadata describes interfaces and functionality through natural language and structured schemas, without specifying any implementation details.
Reasoning from this metadata requires considering possible implementations and identifying plausible ones, which is a task well suited to LLMs' language understanding and pretrained software knowledge.
We therefore implement \mcpsec as a multi-stage LLM-driven system for \nb analysis.


We evaluate \mcpsec on \numSampledTools{} tools from \numMCPs{} widely deployed MCP servers spanning web retrieval, browser automation, SaaS collaboration, and cloud infrastructure.
Ethically conducted PoC testing confirms \numVulTools{} vulnerable tools.
Using only tool metadata, \mcpsec identifies \RQfourMcpsecTP{} of the confirmed vulnerable tools (\RQfourMcpsecRecall{} recall), which outperforms a multi-turn self-refinement agentic baseline that identifies \RQfourAgentTP{} (\RQfourAgentRecall{} recall), at an average analysis cost of \costPerServer{} per MCP server.
Our evaluation also finds that \NoSanitizationNum{} of the examined tools (\NoSanitizationPct{}) pass external data to the LLM context without sanitization, highlighting a broader gap in ecosystem defenses.


\medskip
\noindent
\textit{\textbf{Contributions.}}
This paper makes the following contributions:

\begin{itemize}[leftmargin=*]
    \item We categorize an access-constrained setting as \nb for vulnerability analysis, where the analysts have access to system metadata but cannot inspect the implementation or interact with the deployed system.
    \item We develop a \nb vulnerability analysis framework that speculates about plausible implementations and vulnerability requirements and outputs \emph{Theory-of-Concepts} (ToCs) for later validation when additional access is permitted.
    \item We implement this framework as a prototype called \mcpsec{} and evaluate it on \numSampledTools{} sampled tools from \numMCPs{} widely deployed MCP servers. \mcpsec recovers \RQfourMcpsecTP{} of \numVulTools{} confirmed \IPI vulnerabilities, demonstrating the feasibility of \nb vulnerability analysis in a realistic setting.
\end{itemize}

\medskip
\noindent
In the spirit of open science, we will release all analysis data and source code upon acceptance of this paper.
We will release generated ToCs (and PoCs when available) after the developers of vulnerable MCP servers fix our reported bugs.


%% file: sections/02_background.tex
\section{Background}

\subsection{Existing Vulnerability Analysis Paradigms}
\label{sec:bg-vuln-analysis}


Vulnerability analysis techniques are broadly classified by how much access the analyst has to the target system. 
White-box methods operate on source code or binaries, enabling precise reasoning about program behavior through static analysis, symbolic execution, or formal verification~\cite{wagner2000first, cadar2008klee, shoshitaishvili2016sok, github2026codeql}. 
Gray-box methods such as coverage-guided fuzzing~\cite{demott2007revolutionizing, fioraldi2020afl} require neither full source access nor a behavioral specification, but do require the ability to execute the target and observe coverage feedback. 
Black-box methods~\cite{bau2010state, miller1990empirical} only interact with exposed interfaces and observe responses, requiring no internal system access but still assuming the ability to query the target and receive outputs.

These three paradigms share a common prerequisite: the analyst must observe or interact with the target system, and in doing so, the analyst attempts to speculate about and eliminate false vulnerability hypotheses.

\subsection{Security Requirements Engineering and Threat Modeling}
When access to the target system is limited (the \nb setting), system metadata, such as documentation or interface descriptions, may still provide information about a system's functionality and expected behavior.
Previous work in security requirements engineering and threat modeling provides precedents for reasoning about security properties from such metadata.

In security requirements engineering, descriptions of intended behavior and system goals serve as the basis for constructing abuse cases, misuse cases, and intentional anti-models that describe how an attacker could undermine those goals~\cite{mcdermott_using_1999, van_lamsweerde_elaborating_2004, sindre_eliciting_2005}.
Architectural risk analysis and threat modeling similarly reason from descriptions of system functionality, components, interactions, and trust assumptions~\cite{verdon_risk_2004, myagmar_threat_nodate, tuma_threat_2018}.
STRIDE-based methods, for example, organize threat identification around data flow diagrams and trust boundaries \cite{shostack2014threat}.
These approaches use different forms of metadata to identify potential threats and derive security requirements or mitigations, often during system design~\cite{noauthor_threat_nodate, xiong_threat_2019}.

Our work applies this description-based reasoning to conducting vulnerability analysis on existing targets that have been built or deployed, for which the analysts only have access to the system metadata. 

\subsection{Indirect Prompt Injection Vulnerability}
\label{sec: IPI background}

Prompt injection is a class of rapidly emerging attacks that cause the model in an LLM-based application to treat attacker-controlled content as instructions~\cite{liu_formalizing_2024}.
LLMs are vulnerable to prompt injection attacks because they process instructions and data within a shared context and may follow attacker-supplied instructions that conflict with the application’s intended behavior~\cite{wallace2024hierarchy, zverev_can_2025}.
Prompt injection attacks can be classified as direct or indirect depending on how LLMs pick up the malicious instructions. 
In \emph{direct prompt injection}, the attacker can input malicious instructions directly to the application.
In \emph{indirect prompt injection} (IPI), the attacker embeds these instructions in external data, such as webpages, documents, or tool responses, that the LLM would retrieve and subsequently incorporate into its context~\cite{greshake2023indirect}.
IPI is particularly pervasive because many external data sources that an LLM may fetch are controllable by attackers, and injection payloads are difficult to detect since they often consist of benign natural-language strings that only become malicious when contextualized within the LLM's task~\cite{weng_argus_2026}.
Successful IPI attacks require attacker control over external data and preservation of malicious instructions through application processing into the LLM’s context.

While prior work considered successful exploitation on selected LLMs to establish IPI vulnerability, we argue that this approach provides an unstable basis for vulnerability validation~\cite{yi2023benchmarking, xie_queryipi_2026}. 
Exploitation success can vary substantially with the choice of injection payload and the robustness of the underlying models~\cite{handa2025competencyreasoningopensdoor, li_agentdyn_2026}, but it always requires untrusted data to flow from an attacker-controlled source to a security-sensitive sink.
Thus, in this work, we adopt a vulnerability criterion based on the existence of such a vulnerable data flow, together with additional semantic conditions (detailed in \S\ref{sec:mcp-prelim}).
This criterion helps reduce validation ambiguity arising from payload construction and model selection.


\subsection{Model Context Protocol}
\label{sec:bg-mcp}

The Model Context Protocol (\mcp{}) is a standard interface~\cite{anthropic2024mcp,mcp2024spec} that lets LLM agents invoke external data sources and tools.
A typical \mcp{} server publishes a set of tools.
The metadata of each tool contains its tool name, a natural-language description of functionality, and a list of typed parameters.

A typical MCP interaction proceeds through three phases.
(1)~\textit{Server Registration.}
The user registers an MCP server with the LLM agent host, which then discovers the tool metadata. 
(2)~\textit{Tool Invocation.}
Given a user task, the LLM agent can select from the available MCP tools and invoke the tool on the MCP server.
(3)~\textit{Tool Response.}
The server executes the tool call, which typically involves reading and processing data from external data sources, and returns a free-form result to the LLM host.

%% file: sections/03_nobox.tex
\section{No-Box Vulnerability Analysis}

We conduct vulnerability analysis in the \nb setting through data flow speculation and risk analysis, producing testable attack scenarios that can be validated when additional access becomes available.

\subsection{Vulnerability Analysis as an Information-Centric Framework}
\label{sec:nobox-framework}

Let \(T\) be a target system and let \(\mathcal{M}\) denote metadata about the system's functionality.
Such metadata may include capability specifications, interface and schema declarations, natural-language descriptions, or end-user-facing documentation.
While we assume that \(\mathcal{M}\) truthfully describes the intended functionality of \(T\), it is often a partial and high-level abstraction of the underlying implementation.

We denote by \(\mathcal{I}(\mathcal{M})\) the space of implementations $I$ that implement the functionality described by \(\mathcal{M}\).
\(\mathcal{I}(\mathcal{M})\) contains the true implementation $I^\star$.
Because metadata primarily specifies the \emph{behavior} of the system rather than precisely \emph{how} the behavior is implemented, the size of \(\mathcal{I}(\mathcal{M})\) is effectively unbounded.

Therefore, vulnerability analysis must reason about whether a vulnerability exists across this space of possible implementations and under what conditions it would exist in $I^\star$.
Additional evidence narrows the space and can update our assessment of the likelihood that a vulnerability exists in $I^\star$.
A vulnerability is \emph{necessary} if every implementation consistent with the available evidence contains it and \emph{plausible} if at least one implementation consistent with the available evidence contains it.

Existing vulnerability analysis paradigms differ primarily in the additional evidence they obtain about the unknown implementation.
In a white-box setting, the analyst has access to implementation-level artifacts such as source code, binaries, configuration, or runtime state.
This access narrows the set of compatible implementations, although uncertainty may remain about deployment states or external dependencies.
In a black-box setting, the implementation remains hidden.
Through interaction with $T$, the analyst can observe execution behavior, and each observation can eliminate implementations that cannot produce the observed behavior.
Both forms of access provide evidence that can narrow the space of potential implementations and refine vulnerability
assessments.

Importantly, vulnerability analysis does not necessarily require reconstructing $I^\star$.
Many distinct implementations may share the same vulnerability-relevant properties, such as data flows, transformations, validation mechanisms, and trust-boundary crossings.
Reasoning about these properties can therefore support vulnerability analysis even when the underlying implementation remains uncertain.

\subsection{\NB Vulnerability Analysis}
\label{sec:nobox-nobox}

In the \nb{} setting, the analyst has access only to functionality metadata \(\mathcal{M}\), without implementation artifacts or runtime interaction.
Consequently, the analyst cannot directly determine whether a vulnerability exists in the true implementation \(I^\star\).
Instead, \nb vulnerability analysis seeks to identify vulnerabilities that could arise in implementations consistent with \(\mathcal{M}\), and to characterize the implementation assumptions required for their exploitation.
These objectives are addressed by two complementary analysis processes: data flow speculation and risk analysis. 
Finally, \nb vulnerability analysis outputs concrete exploitation scenarios, each called a \emph{Theory-of-Concept} (ToC), to guide subsequent validation when additional access becomes available.


\subsection{Data Flow Speculation and Risk Analysis}
\label{sec:nobox-dataflow-risk}

Although $\mathcal{M}$ does not include implementation details, it may imply how data must move from one entity to another, which we call an \emph{implied data flow}.
While $\mathcal{M}$ can imply multiple data flows, we further define an \emph{irreducible data flow}: the minimal implementation-agnostic relationship preserved by every implementation consistent with $\mathcal{M}$.

For example, metadata specifying that a user-defined filter selects database records implies that the filter influences a database query.
Its irreducible data flow is
\[
    \textit{user-controlled filter}
    \xrightarrow{\textit{query}}
    \textit{database},
\]
which contains only the essential data entities and transformations required to implement the functionality described by $\mathcal{M}$.
An implementation may realize this relationship through raw query construction, parameterized statements, a restricted parser, or an ORM.
This irreducible data flow makes SQL injection relevant to this example.
However, the existence of an SQL injection vulnerability is contingent on the actual implementation and sanitization used in $I^\star$.

To formulate testable vulnerability hypotheses, we extend irreducible data flows with explicit assumptions about security-relevant implementation choices.
Representing these assumptions as additional data entities or transformations produces \emph{speculated data flows}.
Data flow speculation aims to improve coverage of analysis: it retains plausible flows even when the available metadata cannot establish that they occur in the true implementation.
This \emph{over-approximation} strategy aims to reduce missed vulnerabilities, at the cost of retaining hypotheses that subsequent implementation or runtime evidence may rule out.

However, speculated data flows only suggest that attacker-controlled data can reach a security-sensitive sink.
Whether it arrives in a form that enables exploitation remains an open question, which we address through \emph{risk analysis}, checking each speculated data flow against the necessary conditions for the relevant vulnerability class. 
If at least one metadata-consistent flow satisfies these conditions, it supports a vulnerability hypothesis under the corresponding implementation assumptions.
In the SQL query example, a hypothesized injection is relevant and plausible when a speculated flow incorporates the user-controlled filter into a dynamically constructed query without effective parameterization or sanitization.

\subsection{Theory-of-Concept}
\label{sec:ToC}

Based on both the data flow and the satisfied vulnerability preconditions, we can synthesize an exploitation scenario with additional assumptions, such as attacker roles or possible defenses.
This requires analyzing the amount of attacker control over the source, and answering this question can involve reasoning about \(\mathcal{I}(\mathcal{M})\) (e.g., in terms of access control policy, roles, etc.) or even systems external to \(\mathcal{I}(\mathcal{M})\) (e.g., fetching content from the internet).

The output of \nb vulnerability analysis is what we call a \emph{Theory-of-Concept} (ToC): an exploitation scenario synthesizing speculated data flows and risk analysis results.
A ToC may specify the affected functionality and vulnerability class, the attacker’s role and required capabilities, the implementation-dependent assumptions on which exploitation is based, and the resulting security impact.
Unlike a \emph{Proof-of-Concept} (PoC), a ToC cannot be executed against a concrete implementation.
However, when source code, binaries, or runtime access becomes available, a security analyst can validate the ToC by constructing a PoC.

%% file: sections/04_caseMCP.tex

\begin{figure*}[tb]
  \centering
  \includegraphics[width=\linewidth]{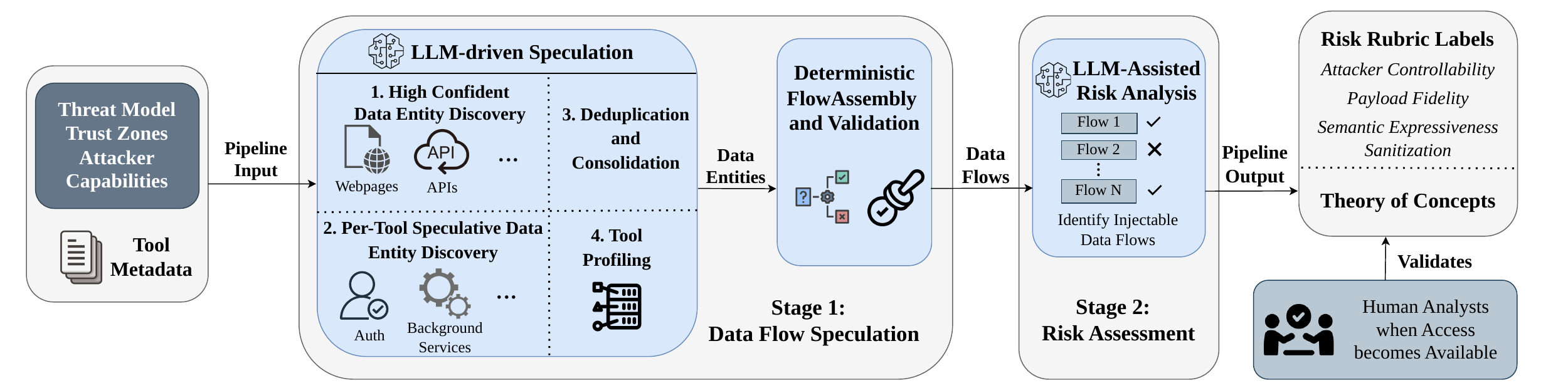}
  \caption{\mcpsec pipeline.}
  \label{fig:pipeline}
\end{figure*}

\section{Finding \IPI Vulnerabilities in MCP Servers}
\label{sec:case-study}

We use MCP servers and \ipi vulnerabilities to demonstrate the necessity and feasibility of \nb vulnerability analysis. 
To measure the prevalence of MCP servers that can only be analyzed in the \nb setting, we enumerated the latest active MCP servers in the official MCP Registry~\cite{mcp_registry}, which contained 18,770 servers at the time of collection.
We classified a server as lacking publicly available implementation artifacts if it declared at least one remote endpoint but provided neither a released package nor a source repository URL.
Under this definition, 3,007 MCP servers (16.0\%) lack such artifacts.
Additionally, 1,210 MCP servers (6.4\%) required users to provide secrets (e.g., API keys) before configuring server connections.
These findings highlight the practical value of \nb{} vulnerability analysis: analysts without access to implementation artifacts or the credentials required for runtime interaction can still formulate testable vulnerability hypotheses from functionality metadata.

We model MCP servers as \nb targets (\S\ref{sec:nobox-nobox}).
The target system $T$ is an MCP server and its tool set.
$\mathcal{M}$ consists of the metadata of the published tools: tool names, natural-language descriptions, and JSON-Schema input specifications.
The MCP specification requires the exposure of these fields to help an LLM select tools and construct valid invocations~\cite{mcp2024spec}.

Next, we discuss the threat model and how to discover \IPI in MCP servers using \nb vulnerability analysis.


\subsection{Threat Model}
\label{sec:threat-model}

An MCP-integrated LLM agent system comprises four relevant entities: a human user, an LLM agent, an MCP server, and the external data sources accessed by its tools.
Table~\ref{tab:trust-zones} summarizes the assumptions assigned to each entity.

\begin{table}[tb]
\centering
\small
\caption{Target system entities, trust zones and threat assumptions.}
\label{tab:trust-zones}
\resizebox{\columnwidth}{!}{
\begin{tabular}{@{}lp{0.78\linewidth}@{}}
\toprule
\textbf{Entity} & \textbf{Trust Zones and Assumptions} \\
\midrule
\textbf{Human User} &
Benign; issues benign tasks. \\
\textbf{LLM Agent} &
Follows the user's intent but may be influenced by instructions embedded in tool responses. \\
\textbf{MCP Server} &
Benign but potentially vulnerable; its code processes external content and may relay it without effective neutralization. \\
\textbf{External Data Sources} &
Untrusted; content can be attacker-controllable. \\
\bottomrule
\end{tabular}
}
\end{table}

We consider a remote third-party attacker who can place instructional content in a source that an MCP tool may subsequently consume.
The attacker may publish public content, contribute content through a legitimate account on an authenticated platform, or influence an upstream artifact that is later retrieved by the user or tool.
Content is considered attacker-controlled if the attacker can influence its contents, even if it is later retrieved and stored locally as part of a cloned repository, installed package, or cached response.
We distinguish an attacker-controlled \emph{source} from the \emph{medium} through which the tool accesses it.
For example, an attacker may publish a web page indexed by a search engine without compromising the search engine itself.

The attacker cannot modify the MCP server, client, or LLMs' weights. 
Direct prompt injection, malicious MCP servers, transport compromise, and local compromise are outside our scope.

The attacker's objective is to introduce a prompt injection payload into the LLM's context while preserving its semantic expressiveness and fidelity.
The corresponding security property is \emph{context integrity}: content controlled by a remote attacker must be mediated so that it cannot be interpreted as instructions to LLMs.
As discussed in \S\ref{sec: IPI background}, we consider an MCP tool vulnerable to \ipi{} if there exists a data flow that carries attacker-controlled content into the LLM's context and satisfies the semantic conditions required for that content to function as an instruction.

\subsection{\NB Analysis of \IPI Vulnerabilities in MCP Servers}
\label{sec:mcp-prelim}

We apply the data flow speculation and risk analysis of \S\ref{sec:nobox-dataflow-risk} to \ipi, starting from its irreducible source-to-sink data flow.
The source set $\mathcal{S}$ contains external origins whose content an MCP tool may consume, and the sink set $\mathcal{K}$ contains the LLM context into which the agent incorporates a tool's response.
For a tool whose metadata entail that externally sourced content contributes to its response, the irreducible data flow is
\[
    s \in \mathcal{S}
    \longrightarrow
    \textit{MCP tool}
    \longrightarrow
    k \in \mathcal{K}.
\]
This relationship indicates that external content can be transmitted to the LLM context, making \ipi vulnerabilities relevant to MCP servers.

Data flow speculation extends the irreducible data flow into candidate speculated data flows by making explicit assumptions about implementation-dependent entities and operations, such as the specific source, intermediate storage, parsing, transformation, field selection, validation, and sanitization.
Each speculated data flow corresponds to the subset of \(\mathcal{I}(\mathcal{M})\) whose implementations satisfy these assumptions.

Risk analysis evaluates each speculative path against four \ipi{}-specific preconditions:
\[
    \Phi_{\mathrm{IPI}}
    = \{\alpha_{\mathrm{fid}},\alpha_{\mathrm{ctrl}},
         \alpha_{\mathrm{sink}},\alpha_{\mathrm{san}}\}.
\]

\textit{1) Payload Fidelity ($\alpha_{fid}$).}
This axis assesses how faithfully attacker-controlled source content is preserved relative to the source when it reaches the LLM's context.
Verbatim format preserves the content, whereas transformations such as summarization and field extraction can alter or discard it.

\textit{2) Attacker Controllability ($\alpha_{ctrl}$).}
This axis collapses two metrics, \emph{Attack Vector} and \emph{Privileges Required}, into a categorical scale suited to MCP. 
Distinguishing public/supply-chain channels~\cite{greshake2023indirect} from authenticated access helps clarify the prerequisites for placing a payload where the tool may consume it.

\textit{3) Semantic Expressiveness ($\alpha_{sink}$).}
This axis measures whether the output representation can express instructions.
Free-form text can convey instructions, whereas strictly constrained outputs, such as fixed-format hashes or booleans, offer limited expressive capacity.

\textit{4) Sanitization ($\alpha_{san}$).}
This axis evaluates the presence of explicit defenses against injected instructions before the tool returns its output, such as instruction-aware filtering (e.g., regex sanitization) or rejection of suspicious content.

A speculative path that coherently satisfies these preconditions yields a metadata-grounded \emph{plausible vulnerability hypothesis}. 
For each such hypothesis, an \ipi{} ToC identifies the affected tool, the attacker-controlled source and access required, the assumed source-to-context path and transformations, the relevant defensive assumptions, and the potential effect on the agent.
The ToC makes the conditions underlying the plausible vulnerability explicit, making it useful for future PoC validation when more access is available.

%% file: sections/05_system_design.tex
\section{\mcpsec: System Design}
\label{sec:system-design}

We develop \mcpsec, a prototype that instantiates the no-box vulnerability analysis framework to analyze MCP servers for \ipi{} vulnerabilities.
As shown in Figure~\ref{fig:pipeline}, \mcpsec performs a two-stage analysis to speculate about \ipi vulnerabilities in the given MCP server.
We use an LLM as the primary reasoning instrument because tool metadata combines natural-language descriptions with structured schemas while leaving implementation details implicit, requiring semantic interpretation and software knowledge beyond schema parsing or fixed rules.
Stage~1 uses the LLM to speculate about the tools' underlying implementations and then applies deterministic flow assembly and structural validation to regulate and organize its outputs into well-formed candidate data flows.
In Stage~2, the LLM assesses each validated flow against the four risk axes defined for \IPI (\S\ref{sec:mcp-prelim}) and, for each flow identified as vulnerable, outputs a ToC describing a concrete attack scenario together with the four risk labels.

\subsection{Stage 1: Data Flow Speculation}
\label{sec:stage1}

Stage~1 takes tool metadata and produces a set of candidate source-to-sink data flows.
To maximize coverage of potential vulnerabilities despite incomplete metadata, \mcpsec prompts the LLM to include unstated entities and edges that remain plausible given the tool's functionality and the threat model, following the over-approximation strategy in \S\ref{sec:nobox-dataflow-risk}.
Stage~1 contains two steps: \emph{LLM-driven Speculation} and \emph{Deterministic Assembly}.

\smallskip
\noindent\textit{\textbf{LLM-driven Speculation.}}
First, entity discovery identifies the data entities.
An initial LLM call extracts entities that can be inferred with high confidence (e.g., external APIs, storage systems, and data pipelines) from all tool metadata of the MCP server, establishing a shared foundation across tools.
Then, per-tool LLM calls propose additional speculative entities whose existence is plausible but not directly stated in the metadata (e.g., intermediate caches, authentication components, and background services).
We instruct the LLM to admit low-confidence entities even from weak or analogical signals because tool metadata can be too concise to be self-explanatory.
We retain such entities to maximize the number of candidate data flows that can be composed.
A consolidation pass merges and deduplicates speculative entities across tools, and the consolidated set of entities forms a \emph{data flow context}.

Tool profiling produces the edges: for each tool, an LLM call receives the data flow context and the tool metadata and produces a \emph{tool profile}.
The tool profile records the LLM's predictions of data flow direction, whether transferred data may contain attacker-controlled content, and the transformations and output fidelity of the tool.
Entity discovery and tool profiling together realize the deliberate over-approximation principle, and their LLM outputs will then be processed in a more deterministic way.

\smallskip
\noindent\textit{\textbf{Deterministic Flow Assembly and Validation.}}
\label{sec:stage1-det}

A deterministic algorithm assembles the predicted entities and tool interactions into candidate data flows and marks potential injection entry points based on the tool profiles.
A structural validator then checks whether, in each assembled flow, injection edges connect valid nodes and a path exists from the attacker-controlled source to the tool's output sink.
These checks constrain the structure of the LLM-generated hypotheses and ensure that malformed data flows cannot silently propagate into downstream reasoning when there is no oracle to detect them.
The resulting data flow graph is internally consistent before it is processed by the LLM for risk assessment.

\subsection{Stage 2: LLM-Assisted Risk Assessment}
\label{sec:stage2}

The plausible data flows generated in Stage~1 support vulnerability hypotheses only if they satisfy the preconditions required for exploitation.
Stage~2 assesses each speculated and validated data flow against the risk assessment rubric, producing risk labels for every flow and a ToC for each flow identified as vulnerable.

\RiskRubricTable{}
\smallskip
\noindent\textit{\textbf{Risk Assessment Rubric.}}\label{sec:risk-rubric}
Intuitively, not all speculated data flows carry equal risk.
A data flow that returns a numeric account balance poses a qualitatively different threat than one that relays verbatim issue comments into the LLM context.
Thus, we define a risk assessment rubric to instantiate the four preconditions of \S\ref{sec:mcp-prelim} for MCP \ipi{}.
The rubric quantifies the risk of a speculated data flow along four independent dimensions: \emph{payload fidelity}, \emph{attacker controllability}, \emph{semantic expressiveness}, and \emph{sanitization}.
Table~\ref{tab:risk-rubric} summarizes the axes and their levels.
The labels allow analysts to prioritize more vulnerable flows, thereby compensating for Stage~1's deliberate over-approximation.

\smallskip
\noindent\textit{\textbf{Producing Outputs.}}\label{sec:mcpsec-outputs}
For each validated data flow, the LLM receives the data flow, the tool metadata, the risk assessment rubric, and the threat-model context (\S\ref{sec:threat-model}) and then analyzes the data flow against the rubric.
It produces two outputs, risk labels for each axis of the \emph{risk assessment rubric} and a ToC, in the same LLM call to bind them to a shared reasoning trace.
The ToC is a short natural-language narrative that walks through a concrete \ipi{} scenario: the attacker model and the external surface under the attacker's control; the planting step that places the payload where the tool will fetch it; and the fidelity with which the tool's output preserves that payload en route to the LLM context.
If the tool has at least one such flow, \mcpsec flags the tool as vulnerable.

As an output of the \mcpsec pipeline, a ToC provides an illustrative attack scenario that helps a downstream analyst test the risk rubric directly against the source code (\S\ref{sec:rq2}) or runtime instrumentation (\S\ref{sec:rq3}).
The per-axis labels provide a more complete view of the predicted risk and enable direct comparison with ground-truth labels during evaluation (\S\ref{sec: eval design}).

\subsection{\mcpsec Implementation}
We implemented \mcpsec as a modular Python prototype comprising approximately 3.4K lines of Python code and 468 lines of fixed prompt text.
It uses the LLM through API calls with structured output for entity speculation, tool profiling, and risk assessment, while deterministic algorithms consolidate entities and assemble and validate data flows.

%% file: sections/06_experiment.tex
\section{Experiment}
\label{sec:experiment}

We outline the dataset construction, experiment setup, and outputs for evaluating \mcpsec.

\subsection{Evaluation Dataset}
\label{sec: dataset}
\EvalTargetTable{}

Our dataset contains \numMCPs{} MCP servers and \numSampledTools{} sampled tools, detailed in Table~\ref{tab:eval-targets}.
We draw the servers randomly from 86 unique, verified servers compiled from the MCP Market top-100 leaderboard, ranked by GitHub stars~\cite{mcpmarket2026}, after removing deleted (3) and archived (1) repositories and deduplicating monorepo entries to avoid over-representation.
Because several servers expose more than 100 tools, we randomly sample 10 tools per server, retaining all tools when fewer are available.
We group the servers by external data interaction pattern, yielding four categories that differ primarily in attacker controllability and source provenance: public web retrieval (A), browser-mediated interaction (B), authenticated SaaS collaboration (C), and infrastructure and cloud platforms (D).

Every server accesses at least one external data source, satisfying the precondition our threat model requires for \ipi{} (\S\ref{sec:threat-model}).
The servers are high-profile: 12 of the \numMCPs{} servers are officially maintained by their service providers, and their stargazer counts (mean \meanStars{}, median \medianStars{}) reflect broad adoption, making them realistic targets on which defenses are more likely to already be in place.
Finally, we select open-source servers so that later stages can ground \mcpsec's predictions in source code and runtime behavior, while withholding that source and those deployments from \mcpsec to faithfully simulate the \nb{} setting.

\subsection{Experiment Setup and Produced Artifacts}

To evaluate the performance of \mcpsec, we consider a single-turn LLM baseline and a multi-turn self-refinement agentic baseline. 

For each target tool, the single-turn baseline makes one API call over the server’s sampled registration metadata.
Its consolidated system prompt preserves \mcpsec's threat model, four-axis risk rubric, and ToC requirements, holding the task specification fixed to share the same understanding of the task with \mcpsec.
The prompt omits entity discovery, DFD reconstruction, deterministic graph assembly, and graph-conditioned reasoning, allowing us to isolate the contribution of \mcpsec's two-stage workflow.

The self-refinement agentic baseline uses the same initial prompt as the single-turn baseline. 
After obtaining the initial response, we provide a fixed self-refinement prompt instructing the LLM to revise its output or terminate when done. 
The details of the agentic baseline setup are outlined in \S\ref{appendix:multi-turn baseline}.
For approximately two-thirds of the tools, the agent determined that no refinement was necessary, suggesting confidence in its initial reasoning.

Since the first turns of both methods are the same (with the same initial prompt), but the agentic baseline additionally allows iterative refinement. We therefore focus our main evaluation on the agentic baseline as the more capable comparison and report the single-turn baseline results in Appendix~\ref{appendix:single-turn LLM results}.

Among the \numSampledTools{} tools, the agent baseline flagged \numAgentBaselineCandidates{}, of which \numBaselineMatchedFlows{} overlap with the \numFlows{} tools flagged by \mcpsec{}.
Results are evaluated against human-annotated labels in \S\ref{sec:evaluation}.

The base LLM in the baselines and \mcpsec is fixed at GPT-5.4, with \texttt{high} reasoning effort.
We also ran \mcpsec with locally hosted Qwen3.8-27B and DeepSeek-V4.1-Flash to investigate model sensitivity in Appendix~\ref{appendix:model-sensitivity}.

%% file: sections/07_evaluation.tex


\section{Evaluation}
\label{sec:evaluation}

\EvalOutlineTable


We previously framed vulnerability analysis as reasoning about possible implementations and making assumptions about underlying vulnerabilities when observation of the target system is imperfect or limited (\S\ref{sec:nobox-framework}).
\NB analysis begins with only metadata and therefore reasons over the full space of metadata-consistent implementations.
Access to metadata, access to source code, and runtime observations progressively increase observability, narrow the possible implementation space, and support correspondingly stronger vulnerability claims.
We organize our evaluation around this progression through four primary research questions, as outlined below.

\subsection{Evaluation Design and Annotation Protocol}
\label{sec: eval design}

While \mcpsec is an automatic pipeline, there exists no automatic oracle for deciding whether a \nb vulnerability hypothesis is consistent with a tool's metadata.
Thus, we rely on human evaluation. 

Two kinds of human labels are required for our evaluation, and they differ in what they are used to judge. 
The first kind judges \mcpsec's outputs (risk labels and ToCs) under progressively stronger evidence (RQ1--RQ3), despite the evaluators transitioning from peers under shared information constraints (the \nb setting) in RQ1 to evidence-informed oracles in RQ2--RQ3.
The second kind judges the tools themselves through controlled PoCs and yields vulnerability labels (RQ4).
Table~\ref{tab:eval-progression} summarizes the evidence available, the resulting labels, and the purpose of each stage.

\textbf{Staged evaluation of risk assumptions and ToCs (RQ1--RQ3).}
Evaluators annotate each of the \numFlows{} tools flagged by \mcpsec in three stages, each adding one form of evidence.
These tools include \numAgentBaselineMatchedFlows{} of the \numAgentBaselineCandidates{} tools flagged by the agent baseline.

Risk-rubric labels are independently assigned by evaluators based solely on the evidence available in RQ1 and RQ2, without access to \mcpsec's outputs.
This helps to measure whether \mcpsec's risk labels are aligned with human judgment, especially in RQ1 where both human evaluators and \mcpsec are in the \nb setting.
In addition to assigning risk labels, evaluators use different levels of evidence to judge the plausibility of the predicted ToC in each stage.
In RQ1, evaluators see only the tool metadata and judge whether the predicted injection source and ToC are plausible.
For RQ2, evaluators are additionally given the source code of each tool and asked whether the predictions in the \nb setting remain plausible when evaluated against static analysis.
For RQ3, we additionally allow evaluators to execute target tools and observe whether the predicted vulnerable flow manifests at runtime and with what payload fidelity.
RQ3 uses benign tool parameters and validates runtime reachability rather than exploitability, which is evaluated separately in RQ4.

\textbf{Vulnerability labeling (RQ4).}
RQ1--RQ3 assess only the tools that \mcpsec flags as vulnerable.
To quantitatively measure the vulnerability recovery performance of \mcpsec, we establish ground-truth vulnerability labels for our dataset (\numSampledTools{} tools) by constructing PoCs.
Because some target MCP servers interact with commercial services, we construct PoCs only for tools that can be safely and ethically tested in a controlled setting.
We describe the details of this procedure in Appendix~\ref{appendix:rq4-poc} and discuss the ethical considerations there.
We label \numVulTools{} tools as vulnerable to \ipi{} attacks; the remaining tools are excluded because they either cannot be confirmed as vulnerable or are unsuitable for PoC construction.
We then design RQ4 to compare the vulnerability discovery recall of \mcpsec and the baseline against these labels.

\textbf{Annotation quality.}
The annotations are performed by a subset of the authors, all with experience in data flow analysis and security research.
To assess consistency, at least two evaluators annotated each tool.
Before evaluators resolved label disagreements, exact agreement across all paired categorical judgments for RQ1--RQ3 was \humanAgreementOverallPct{}.
After calculating agreement scores, each evaluator pair reviewed the disagreements and then recorded a consensus label and rationale; unresolved cases were referred to a third evaluator.

\subsection{RQ1: Are \mcpsec-Generated ToCs and Risk Labels Plausible from Metadata Alone?}
\label{sec:rq1}

\noindent\textit{\textbf{Evaluation target.}} 
In RQ1, we assess whether the risk hypotheses and ToCs generated by \mcpsec are internally consistent and grounded in the available tool metadata.
For each plausible vulnerability candidate identified by \mcpsec, the evaluators observe only the tool metadata and evaluate the generated artifacts along six dimensions: (1) whether the identified external data source is plausible for the tool, (2--5) whether the tool metadata supports each of the four risk-rubric labels (\S\ref{sec:risk-rubric}), and (6) whether the ToC is coherent and plausible. 
RQ1 results are shown in Table~\ref{tab:rq1-results}. 

\smallskip 
\noindent\textit{\textbf{Result Analysis}}
\RQOneResultsTable
Evaluators find that \mcpsec identifies a plausible external data source in \stageOnePlausiblePct{} of its \numFlows{} flagged tools and produces a plausible ToC in \stageOneTocCoherencePct{}.
The decrease from source plausibility to ToC plausibility suggests that identifying a plausible external source does not always lead to a plausible end-to-end attack hypothesis, particularly when tool metadata are concise.

Beyond source and ToC plausibility, \mcpsec achieves \stageOneMeanAgreementPct{} agreement on average with human judgments across the four risk-rubric axes.
The highest-agreement axes are \texttt{Semantic\_Expressiveness} (\stageOneSemanticExecPct{}) and \texttt{Attacker\_Controllability} (\stageOneAttackPreqPct{}), suggesting that output form and attacker access are often directly inferable from tool descriptions.
The lowest-agreement axis is \texttt{Sanitization} (\stageOneSanitizationPct{}), for which most disagreements occur when \mcpsec labels sanitization as \texttt{partial} but evaluators label it as \texttt{none} (\sanitizationDisagreementNum{} cases).
This pattern suggests a difference in how human evaluators and \mcpsec interpret what constitutes partial sanitization from the available metadata, particularly whether certain transformations or processing steps provide meaningful sanitization against the hypothesized injection.
For \texttt{Attacker\_Controllability} (\stageOneAttackPreqPct{}), \mcpsec most often labels flows as \texttt{authenticated\_remote}, whereas evaluators downgrade them to \texttt{implausible} (\downgradedAttackPrereq{} cases).
These disagreements occur more often for write-only tools for which evaluators find no plausible return-path injection surface.

On the \numAgentBaselineMatchedFlows{} tools flagged by both \mcpsec and the agentic baseline, \mcpsec achieves \stageOneAgentMatchedMeanAgreementPct{} agreement on average across the four risk-rubric axes, compared with \agentBaselineMeanAgreementPct{} for the baseline.
The gap comes from \texttt{Payload\_Fidelity} (\stageOneAgentMatchedDataFidelityPct{} vs.\ \agentBaselineDataFidelityPct{}) and \texttt{Sanitization} (\stageOneAgentMatchedSanitizationPct{} vs.\ \agentBaselineSanitizationPct{}), the two axes that require tracing data from the source to the tool output.
Although the baseline receives the same metadata together with offline tools and self-refinement, it does not match \mcpsec's explicit data flow reconstruction on these axes.

\smallskip 
\noindent\textit{\textbf{Conclusion.}}
In the \nb setting, \mcpsec can construct ToCs that are largely plausible and internally consistent under human evaluation.
Agreement is weaker for properties that depend on implementation details not directly exposed by metadata, most notably the prediction of sanitization.
Overall, RQ1 measures the \emph{metadata-grounded plausibility} of \mcpsec's analysis but not whether its hypotheses accurately characterize the underlying implementation; we examine this distinction using source-code evidence in RQ2.

\subsection{RQ2: Are \mcpsec-Generated ToCs and Risk Labels Plausible Under Source-Code Evidence?}\label{sec:rq2}

In RQ1, both \mcpsec and human evaluators operate under the \nb setting, so evaluation measures agreement rather than factual correctness.
In RQ2, evaluators inspect the source code to examine predicted external data sources and reassess ToCs and risk labels against implementation evidence.
They retain the RQ1 judgments for \texttt{Attacker\_Controllability}, for which the source code does not provide additional evidence, and reassess the other three risk-rubric labels.
These judgments yield an overall ToC verdict: \emph{Plausible}, \emph{Partially Plausible}, or \emph{Implausible}.
RQ2 results are shown in Table~\ref{tab:rq2-results}. 

\smallskip
\noindent\textit{\textbf{Results Analysis.}}
\RQTwoResultsTable
Source-code validation shows that \combinedFeasiblePartialFeasibleToCPct{} of the predicted ToCs remain plausible or partially plausible in the actual implementations: \FeasibleToCPct{} are judged \emph{Plausible}, \PartialFeasibleToCPct{} are judged \emph{Partially Plausible}, and \UnfeasibleToCPct{} (\UnfeasibleToCNum{}) are judged \emph{Implausible}.
Human evaluators also determined that \mcpsec's predictions of external data sources are correct and supported by the source code for \PlausibleSourcePct{} of tools.

To understand why \mcpsec-predicted ToCs do not fully survive implementation validation, we examine the \combinedUnfeasiblePartialFeasibleToCNum{} flows rated \emph{Implausible} or \emph{Partially Plausible}.
A tool may exhibit multiple implementation-level factors: \emph{data not preserved} (\FactorDataNotPreservedNum{}/\combinedUnfeasiblePartialFeasibleToCNum{}, \FactorDataNotPreservedPct{}), where external content is reduced to identifiers, counts, or status codes; \emph{output transformed} (\FactorOutputTransformedNum{}/\combinedUnfeasiblePartialFeasibleToCNum{}, \FactorOutputTransformedPct{}), where summarization or reformatting prevents payload preservation; \emph{effective sanitization} (\FactorEffectiveSanitizationNum{}/\combinedUnfeasiblePartialFeasibleToCNum{}, \FactorEffectiveSanitizationPct{}); and \emph{wrong source} (\FactorWrongSourceNum{}/\combinedUnfeasiblePartialFeasibleToCNum{}, \FactorWrongSourcePct{}).
The dominant failure mode is therefore incidental data loss rather than active security defense: benign transformations performed as part of the tool's intended functionality alter or discard the attack payload before it reaches the sink.
Incorrect source identification also remains a nontrivial factor.
Among the \ImplausibleSourceNum{} incorrect source identifications, \ImplausibleSourceUnfeasibleToCNum{} result in \emph{Implausible} ToCs, indicating that source-identification errors directly contribute to implausible ToCs.

Source-code evidence also lowers agreement on the predicted risk labels.
The decline concentrates on \texttt{Semantic\_Expressiveness} (\stageOneSemanticExecPct{} in RQ1 to \stageTwoSemanticExecPct{} in RQ2) and \texttt{Payload\_Fidelity} (\stageOneDataFidelityPct{} to \stageTwoDataFidelityPct{}).
MCP tool metadata often specify only the functionality of a tool. These two axes depend on what the tool returns and in what form, so metadata-consistent implementations can differ considerably in these two properties.
Since \mcpsec's labels remain fixed across both stages, the decline in agreement reflects evaluators revising their judgments based on additional source-code evidence.
This observation is consistent with our information-centric framework (\S\ref{sec:nobox-framework}), where additional evidence narrows the space of candidate implementations and refines vulnerability assessments.

Furthermore, we observe that explicit defenses are uncommon but effective.
\NoSanitizationPct{} (\NoSanitizationNum{}) of tools apply no sanitization, \PartialSanitizationPct{} (\PartialSanitizationNum{}) apply partial filtering, and only \EffectiveSanitizationPct{} (\EffectiveSanitizationNum{}) apply effective sanitization. 
Effective sanitization includes regex-based filtering and intentional minimization of returned data. 
All \EffectiveSanitizationNum{} of these effectively sanitized tools receive \emph{Implausible} verdicts, which reflects the scarcity of defense mechanisms against prompt injection payloads rather than their ineffectiveness.

On the \numAgentBaselineMatchedFlows{} tools shared with the agentic baseline, \mcpsec retains \stageTwoAgentMatchedMeanAgreementPct{} mean agreement with source-code-grounded judgments, compared with \agentBaselineStageTwoMeanAgreementPct{} for the baseline.
As in RQ1, the gap between \mcpsec and the agentic baseline is concentrated on \texttt{Payload\_Fidelity} and \texttt{Sanitization}.
With the reference labels now grounded in source code, this gap shows that \mcpsec's explicit data flow reconstruction is not only closer to human metadata-based judgment (RQ1) but also more often accurate about how the implementation preserves or filters external content.

\smallskip 
\noindent\textit{\textbf{Conclusion.}}
Source-code evidence largely confirms \mcpsec's \nb outputs, with the majority of ToCs remaining at least partially plausible.
Where ToCs are deemed implausible, the cause is rarely active defense but rather incidental data loss.
This suggests that current MCP servers lack defenses designed specifically to counter indirect prompt injection and that their limited resilience often arises from functional design rather than deliberate security precautions. 
\mcpsec is able to capture this pattern and translate it into plausible ToCs.

\subsection{RQ3: Do ToC-Identified Data Flows Reach the LLM Context at Runtime?}\label{sec:rq3}

\noindent\textit{\textbf{Evaluation target.}} 
RQ1 and RQ2 establish that \mcpsec's hypotheses are largely plausible from metadata and survive source-code scrutiny.
RQ3 tests whether external content reaches tool outputs at runtime in a form that preserves potential injection content.
For each tool, we log data immediately after the remote API call that retrieves external content (\emph{source}) and at the return statement (\emph{sink}), using benign parameters constructed by evaluators to retrieve realistic data.
We assess each flow along two dimensions: the amount of external data reaching the output (\emph{reachability}: complete, partial, metadata-only, or none) and the observed \emph{output form} (verbatim or derived).
Table~\ref{tab:verdict-matrix} maps the cross-product to a verdict label, and Table~\ref{tab:rq3-results} shows RQ3 results.

\begin{table}[tb]
\centering
\small
\caption{Verdict derivation for RQ3.}
\begin{tabular}{@{}lll@{}}
\toprule
\textbf{Reachability} & \textbf{Output form} & \textbf{Verdict} \\
\midrule
complete   & verbatim text & \textbf{Reachable-Verbatim} \\
complete/partial & derived & \textbf{Reachable-Modified} \\
metadata only & \emph{(any)} & \textbf{Reachable-Metadata} \\
none       & \emph{(any)} & \textbf{Not Reachable} \\
\bottomrule
\end{tabular}
\label{tab:verdict-matrix}
\end{table}

\smallskip
\noindent\textit{\textbf{Results Analysis.}}
\RQThreeResultsTable
Across the \numFlows{} tools \mcpsec identifies, \ReachableVerbatimPct{} (\ReachableVerbatimNum{}) are classified as Reachable-Verbatim, \ReachableModifiedPct{} (\ReachableModifiedNum{}) as Reachable-Modified, and \ReachableMetadataPct{} (\ReachableMetadataNum{}) as Reachable-Metadata; the remaining \NotReachablePct{} (\NotReachableNum{}) are classified as Not Reachable.
In total, \TotalInstructionReachablePct{} (\TotalInstructionReachableNum{}/\numFlows{}) of flows can carry attacker-controllable data to the LLM context, either verbatim (\ReachableVerbatimNum{}) or through structural transformations such as JSON reformatting and field selection that preserve injection-capable content (\ReachableModifiedNum{}).
Not Reachable tools are more likely to be write-only or UI-interaction tools, such as the \emph{click} tool of Chromium DevTools MCP, that return only input-derived confirmations or error responses.

\smallskip
\noindent\textit{\textbf{Validating \mcpsec's two-stage design.}}
Cross-referencing RQ2 with RQ3 tests the two stages of \mcpsec separately.
The data flow speculation stage (\S\ref{sec:stage1}) predicts that a flow exists, and runtime reachability validates this prediction.
Of the \FeasibleToCNum{} tools with \emph{Plausible} ToCs, \FeasibleReachableNum{} (\FeasibleReachablePct{}) are reachable at runtime.
Even among the \UnfeasibleToCNum{} tools with \emph{Implausible} ToCs, \NotFeasibleButReachableNum{} (\NotFeasibleButReachablePct{}) are reachable: the speculated flow exists, but it carries only metadata or caller-provided content rather than an attacker-controlled payload.
The flows \mcpsec infers from metadata largely exist in the implementation, but their existence and reachability alone cannot classify a tool as vulnerable.

The risk assessment stage (\S\ref{sec:stage2}) aims to distinguish between reachable data flows to find potentially vulnerable ones by evaluating their vulnerability preconditions against the risk rubric (Table~\ref{tab:risk-rubric}).
To evaluate this design, we investigate how \mcpsec's risk labels (inferred in the \nb setting) correlate with evaluators' RQ2 plausibility verdicts among tools confirmed reachable in RQ3.
We observe that \mcpsec assigns low \emph{payload fidelity} to \NotFeasibleReachableLowFidelityPct{} of the \emph{Implausible} tools but only \FeasibleReachableLowFidelityPct{} of the \emph{Plausible} tools, and low \emph{semantic expressiveness} to \NotFeasibleReachableLowExecPct{} versus \FeasibleReachableLowExecPct{}.
These results suggest that the design of the risk assessment stage is effective in filtering out reachable but non-vulnerable flows, and that \mcpsec's risk labels can contribute to the generation of plausible vulnerability hypotheses.

\smallskip
\noindent\textit{\textbf{Conclusion.}}
Overall, RQ3 confirms that most data flows hypothesized by \mcpsec can carry injection-capable content to the LLM context at runtime. Moreover, the results demonstrate the importance of combining data-flow speculation with risk assessment, as \mcpsec can anticipate exploitability constraints that reachability alone cannot capture.
Since RQ3 only establishes reachability rather than end-to-end exploitability, RQ4 next examines whether \mcpsec recovers ground-truth vulnerabilities.

\subsection{RQ4: Does \mcpsec{} Recover Confirmed Prompt-Injection Vulnerabilities?}
\label{sec:rq4}

\noindent\textit{\textbf{Evaluation target.}}
RQ1--RQ3 assess \mcpsec's vulnerability analysis on the \numFlows{} tools it identified as vulnerable under progressively stronger evidence.

Following \S\ref{sec: IPI background}, we consider that a tool is vulnerable to \ipi{} when a vulnerable data flow exists, without requiring successful exploitation on selected LLMs.
This reduces ambiguity in vulnerability validation caused by the choice of payload and model. 
Under this criterion, we label \numVulTools{} tools as vulnerable and measure how many of them \mcpsec can recover.
As discussed earlier in \S\ref{sec: eval design}, the remaining \RQfourUnlabeled{} tools are not established negatives.
A tool goes unconfirmed either because we cannot rule out the possibility of it being vulnerable or because no PoC could be attempted against it because of ethical considerations or practical constraints.
As in AgentFuzz~\cite{guo2025agentfuzz}, we report recall over confirmed vulnerabilities as the primary metric for RQ4, since unconfirmed tools cannot be reliably treated as negatives.
Results are shown in Table~\ref{tab:rq4-results}.

\smallskip
\noindent\textit{\textbf{Results.}} \mcpsec{} recovers \RQfourMcpsecTP{} of the \numVulTools{} confirmed vulnerabilities (\RQfourMcpsecRecall{}), whereas the agentic baseline recovers \RQfourAgentTP{} (\RQfourAgentRecall{}).
With only one missed vulnerability, \mcpsec demonstrates that it can effectively discover confirmed \ipi{} vulnerabilities in MCP servers.
The agentic baseline also performs strongly, indicating that \nb{} vulnerability analysis remains effective even without explicit data flow reconstruction.

However, prompting-based reasoning alone is insufficient to achieve the same coverage as \mcpsec's structured analysis.
The agentic baseline recovers one more vulnerability than the single-turn baseline (Appendix~\ref{appendix:single-turn LLM results}) given more self-refinement steps, and its misses (\RQfourAgentFNCollab{}/\RQfourAgentFN{}) are concentrated in Category~C (Table~\ref{tab:eval-targets}).
The cause of these misses is methodological.
The agentic baseline evaluates tools holistically, dismissing operations such as merging pull requests or retrieving user profiles as trusted platform interactions.
In contrast, \mcpsec's data flow speculation examines individual response fields, tracing their origins and attacker controllability.
This exposes injection surfaces such as pull request titles and profile biographies that other GitHub users can modify, even when the overall operation appears benign.

\smallskip
\noindent\textit{\textbf{Conclusion.}}
Together with the agentic baseline's strong recall, the RQ4 results demonstrate that \nb{} analysis can recover real \ipi{} vulnerabilities without source-code access or target interaction, while \mcpsec's structured, over-approximating analysis substantially reduces missed vulnerabilities.

\RQFourSelfContainedTable

\subsection{Case Studies}
\label{sec:case-studies}

We highlight two representative cases from our evaluation that illustrate the strengths of \mcpsec's analysis and the limitations of black-box testing.
The first shows a metadata-derived ToC that leads to a working exploit; the second shows a plausible ToC that an independent analyst cannot validate through black-box testing without privileged access.

\textbf{From page text to bidirectional exfiltration channel.}
\mcpsec flags \texttt{evaluate\_script} in \texttt{chrome-\allowbreak devtools-mcp} by identifying a flow from attacker-controlled page content to JavaScript execution that can read same-origin browser state and exfiltrate it to a remote endpoint.
We implement a PoC based on this ToC using the unmodified MCP server and a locally hosted LLM agent, seeding credentials into \texttt{localStorage} and cookies to simulate a prior SSO session.
When asked to ``summarize this page,'' the agent follows an injected instruction and invokes \texttt{evaluate\_script} with a supplied function that sends \texttt{localStorage} and \texttt{document.cookie} to a remote server via an authenticated POST.
This case validates the claim that a ToC produced by \mcpsec can be utilized by human analysts to produce an effective PoC.
Also, it demonstrates the worst-case impact: injected page content can exploit Chromium's network privileges to access and exfiltrate persistent browser data.

\textbf{Black-box testing requires privileged access.}
\mcpsec produces a ToC for \texttt{slack/usergroups\_\allowbreak update} based on a flow in which an attacker first places instructions in the description of a Slack user group.
If another user later updates a different group property, such as its name or handle, the tool may return the pre-existing description to the LLM along with the updated group metadata.
While our source-code inspection supports this ToC, testing this path requires more access.
Slack user groups are available only in paid workspaces, and the endpoint requires an OAuth token with the \texttt{usergroups:write} scope and a workspace role allowed to manage user groups\footnote{\url{https://docs.slack.dev/reference/methods/usergroups.update}}.
Without this access, a black-box test cannot obtain a successful tool response.
By contrast, \mcpsec can still identify this conditional flow from metadata and record the assumptions that a later authorized test must check.
For such permission-gated tools, a metadata-derived hypothesis may be the strongest result available to an independent analyst.

%% file: sections/08_discussion.tex

\section{Discussion}
\label{sec:discussion}

\subsection{Practical Role and Applicability of No-Box}

Our results demonstrate that the generation of vulnerability hypotheses does not always require implementation access.
Across 20 widely deployed MCP servers, our proposed method can construct reasonable IPI vulnerability hypotheses and achieve a high recall rate when validated against ground truth.
The analysis cost only \$1.65 per server (when using GPT-5.4) on average, required no credentials or source code, and did not involve the risk of disturbing live services.
Thus, we prove that \nb is not only feasible but also cost-effective, filling the gap between the existing vulnerability analysis paradigms.

At MCP ecosystem scale, \nb analysis can serve as a cost-effective and light-weight method at the point of server listing or installation, with minimal deployment or configuration overhead. 
The produced ToCs can then be used to prioritize further white-box, gray-box, or runtime validation, or to warn users and hosts about potential vulnerabilities.

While extending \nb analysis to more target systems and vulnerability classes can be worth investigating, its feasibility and performance may depend on two dimensions. 
Along the \emph{target system} dimension, the observable metadata need to be semantically meaningful for analysts to infer an irreducible data flow.
Software ecosystems such as browser extensions, CI/CD actions, and serverless applications, whose metadata expose entry points, data interfaces, and privileged capabilities, can also be considered suitable targets.
Across different \emph{vulnerability classes}, the approach works best when exploitation conditions can be defined using metadata alone. 
This makes SQL injection, XSS, SSRF, and path traversal highly suitable. 
However, memory corruption and race conditions are less suitable, as their key triggering conditions are deeply obscured by low-level implementation details.

\subsection{A Conceptual Gap in MCP Security}
Many of the PoCs constructed in RQ4 required only placing a fixed injection payload in an attacker-controllable external source, rather than designing a sophisticated payload or attack chain.
Our analysis suggests that such straightforward attacks remain possible because sanitization is largely absent from the path between external data and the LLM context: \NoSanitizationPct{} of the examined tools apply no sanitization along this path.
These implementations appear to treat a valid API response as safe model input, placing the trust boundary at the API response rather than at the external data source.
This mismatch helps explain why simple injection attacks remain viable across our dataset.

Developing safer MCP server implementations should begin with output minimization.
When a task requires only an identifier, count, or status, returning external free-form content creates an unnecessary injection surface and should be avoided.
When such content is necessary, servers should preserve its provenance and separate it structurally from server-generated output so that hosts can distinguish data from instructions~\cite{hines2024spotlighting,debenedetti2025camel}.
Heuristic sanitization methods such as length limits and rule-based filters provide only partial protection because short or adaptive payloads may bypass them; additional LLM guardrails should be considered to support defense in depth~\cite{zhan2025adaptive,jacob2025promptshield}.
MCP developers should therefore treat external content as the trust boundary and be more mindful of the security implications in their implementations.

\subsection{Limitations}
\label{sec:limitations}

We attribute part of the high recall performance of MCPSEC on IPI vulnerability analysis to the fact that MCP registration metadata are standardized (by the MCP specification \cite{mcp2024spec}) and semantically informative.
In target systems with poorer specifications or weaker interface conventions, the same method may produce substantially more false positives or fail to generate useful hypotheses. 

Our evaluation may not fully represent the broader MCP ecosystem or its deployed implementations. 
We selected open-source servers to enable subsequent validation of \mcpsec’s predictions, and prioritized high-profile servers as realistic targets where defenses are more likely to be in place.
This selection may under-represent private or less popular servers with different security characteristics.

\mcpsec reasons about each tool in isolation and models a single external-source-to-tool-to-context path, leaving multi-hop attack chains, host-side prompt construction, cross-turn memory accumulation, and inter-server compositions out of scope. 
LLM-based data flow speculation also remains non-deterministic.
The deterministic assembly step and repair loops substantially constrain model behavior but do not eliminate variance entirely. 


%% file: sections/09_related_work.tex


\section{Related Work}
\label{sec:related-work}

\subsection{No-Box Vulnerability Analysis}

The implementation or execution access required by conventional vulnerability analysis (\S\ref{sec:bg-vuln-analysis}) is unavailable in the \nb setting.

Security requirements engineering and threat modeling provide precedents for reasoning about security from system descriptions~\cite{mcdermott_using_1999,sindre_eliciting_2005,verdon_risk_2004,tuma_threat_2018}.
For example, van Lamsweerde~\cite{van_lamsweerde_elaborating_2004} derives potential attacks from system goals and attacker capabilities to elicit security requirements.
Our work applies description-based reasoning to vulnerability analysis of existing systems: externally available metadata supports testable vulnerability hypotheses whose implementation assumptions can be checked when additional access becomes available.

Prior work also uses \emph{\nb} to describe adversarial attacks against neural networks and LLM-based agents with limited knowledge of or access to the target~\cite{li_practical_2020,zhang_practical_2025,shi_prompt_nodate}.
These studies focus on attack construction rather than on deriving vulnerability hypotheses from functionality metadata.

To the best of our knowledge, ours is the first work to conduct and evaluate vulnerability analysis under this metadata-only setting.

\subsection{Indirect Prompt Injection in LLM-Based Systems}
Prior studies demonstrate \IPI through retrieved content and multi-modal inputs~\cite{greshake2023indirect,bagdasaryan2023abusing}, including attacks that induce unintended tool use or data exfiltration in LLM agents~\cite{zhan2024injecagent,debenedetti2024agentdojo}.
Defenses include runtime detection~\cite{jacob2025promptshield,zhu2025melon} and instruction-data separation~\cite{hines2024spotlighting,chen2025struq,wallace2024hierarchy,debenedetti2025camel}, although adaptive attacks can bypass these mechanisms~\cite{zhan2025adaptive}.

Research on pre-deployment vulnerability detection for LLM agents remains limited compared with research on post-deployment vulnerability analysis~\cite{liu2024llmsmith, guo2025agentfuzz, agentarmor2025}.
LLMSmith~\cite{liu2024llmsmith} extracts call chains to code-execution sinks from framework source code and confirms them by prompting deployed apps.
AgentFuzz~\cite{guo2025agentfuzz} applies directed greybox fuzzing to find taint-style vulnerabilities in LLM agents and relies on feedback from live runs.
AgentArmor~\cite{agentarmor2025} defends against prompt injection by building program dependence graphs from the runtime traces of agents.
Such access may be unavailable, for example, for closed-source MCP servers or servers that require privileged access.
Designed for the \nb setting, \mcpsec requires no such access and supports analysis both before deployment and in deployed environments where access is unavailable.


%% file: sections/10_conclusion.tex

\section{Conclusion}
\label{sec:conclusion}

In this work, we conduct and evaluate vulnerability analysis in the \nb setting, where analysts reason about vulnerabilities from metadata alone when source code, binaries, and runtime interaction are unavailable.
We instantiate it as \mcpsec, which speculates about data flows from MCP tool metadata, assesses their risk, and produces ToCs as testable hypotheses of \ipi{} vulnerabilities.
In our evaluation on \numMCPs{} widely deployed MCP servers, \mcpsec's labels and ToCs are largely consistent with those of human evaluators and hold under source-code and runtime validation.
It recovers \RQfourMcpsecTP{} of \numVulTools{} confirmed vulnerabilities with \RQfourMcpsecRecall{} recall, compared with \RQfourAgentRecall{} for an agentic baseline.
\NB{} analysis can therefore serve as a low-cost triage layer that directs later white-box or black-box validation, and we hope it motivates future work on other vulnerability classes and target systems.

%% file: sections/appendix/ethics.tex
\section{Ethical Considerations}
\label{sec:ethical-considerations}

The primary ethical concern in this study is exposing third-party users to prompt-injection payloads while constructing ground-truth vulnerability labels.
An end-to-end PoC would normally require us to place a prompt-injection payload in an external data source, where other users might encounter it.
We therefore separate the attacker-controllability test from payload reachability testing.
For sources under our control, we place a semantically inert canary in the remote resource and remove it after testing.
Only after retrieving the canary do we replace it with prompt-injection text inside the locally running MCP server process.
The prompt-injection text is never stored on a third-party service.
We exclude tools when testing would require destructive or irreversible operations, affect resources outside our control, expose adversarial content to other users, or rely on permissions unavailable to our research accounts.
Appendix~\ref{appendix:rq4-poc} describes this protocol in detail.
These safeguards limit the tools we can test directly.
\NB{} vulnerability analysis becomes particularly useful in such cases because it can identify potential vulnerabilities when system access is unavailable or direct testing would be unsafe.

We ran the case-study exploit in an isolated environment with a locally hosted webpage, a local LLM agent and MCP server, a controlled browser instance, and a researcher-controlled collection endpoint.
All credentials and browser state used in the experiment were synthetic.
We used no real user credentials or private data.

We reported the case-study finding to the relevant maintainer.
At the time of writing, coordinated disclosure of actionable RQ4 findings is ongoing.

%% file: sections/appendix/model_sensitivity.tex
\section{Sensitivity to the Base Model}
\label{appendix:model-sensitivity}

To assess sensitivity to the base model, we rerun \mcpsec{} on the same \numSampledTools{} tools with locally hosted Qwen3.8-27B (\texttt{xhigh} reasoning effort) and DeepSeek-V4.1-Flash (\texttt{high} effort), keeping the metadata and prompts fixed.
We reuse the consensus risk labels from RQ1 and the payload-delivery labels from RQ4; RQ2 and RQ3 are not repeated.

Table~\ref{tab:model-sensitivity} shows that both alternative models recover all \numVulTools{} tools with confirmed instruction-bearing payload delivery, while flagging more candidates than GPT-5.4.
On the same 125 tools, mean agreement across the four risk-rubric axes is 83.8\% for GPT-5.4, 74.8\% for Qwen, and 82.2\% for DeepSeek.
Thus, vulnerability recovery remains high across the tested models, while agreement with human risk judgments varies with the base model.

\begin{table}[!ht]
\centering
\small
\setlength{\tabcolsep}{4pt}
\caption{Base-model sensitivity. Recovery uses RQ4 payload-delivery labels. Risk-rubric agreement uses RQ1 consensus labels on uniquely matched tools; the final rows use the same 125 tools for all models.}
\label{tab:model-sensitivity}
\resizebox{\columnwidth}{!}{%
\begin{tabular}{@{}lccc@{}}
\toprule
\textbf{Measure} & GPT-5.4 & Qwen3.8-27B & DeepSeek Flash \\
\midrule
\multicolumn{4}{@{}l}{\textit{Vulnerability recovery}} \\
Flagged tools / 177 & 143 & 154 & 164 \\
Recovered / 95 & 94 & 95 & 95 \\
Missed / 95 & 1 & 0 & 0 \\
Recall & 98.9\% & 100.0\% & 100.0\% \\
\midrule
\multicolumn{4}{@{}l}{\textit{Agreement with RQ1 human labels}} \\
Matched tools / 143 & 143 & 142 & 126 \\
Payload fidelity & \stageOneDataFidelityPct & 79.6\% & 79.4\% \\
Semantic expressiveness & \stageOneSemanticExecPct & 83.1\% & 81.7\% \\
Attacker controllability & \stageOneAttackPreqPct & 85.9\% & 74.6\% \\
Sanitization & \stageOneSanitizationPct & 52.1\% & 92.9\% \\
Mean (four axes) & \stageOneMeanAgreementPct & 75.2\% & 82.1\% \\
\midrule
\multicolumn{4}{@{}l}{\textit{Agreement on the common RQ1 cohort}} \\
Common tools / 143 & 125 & 125 & 125 \\
Mean (four axes) & 83.8\% & 74.8\% & 82.2\% \\
\bottomrule
\end{tabular}%
}
\end{table}
\FloatBarrier

%% file: sections/appendix/baselines.tex
\section{Baseline Details}
\label{appendix:baselines}

\subsection{Agentic Baseline Setup}
\label{appendix:multi-turn baseline}

For the multi-turn self-refinement baseline, we use the OpenAI Agents SDK.
The agent is placed in a Docker container with access to the shell and file-system tools.
Network access is disabled, and the agent cannot access the target MCP server or its source.
The workspace holds only \emph{toolset.json} (the server’s sampled registration
metadata) and \emph{task.md} (the user prompt).

After the agent first outputs its vulnerability verdict and risk rubric labels, we send a fixed self-refinement message:
\emph{``You now have another chance to revise your answer if you think it is necessary. If you think there is no point of continuing refining the results for this task, output DONE.''}
The loop ends whenever \emph{``DONE''} is detected in the output.
Across the \numSampledTools{} tools, we observe that the agent performs no refinement in 116 cases and 1 refinement in 58 cases. The maximum number of refinement turns is 5.
Additionally, we noticed that refinement revised the analysis 61 times and changed the vulnerability verdict or risk labels for 15 tools.
The agent made only 9 shell/file tool calls in total.

\subsection{Single-Turn Baseline Results}
\label{appendix:single-turn LLM results}

Table~\ref{tab:baselines} compares \mcpsec{} with both baselines.
RQ1 and RQ2 agreement uses the \numSharedBaselineFlows{} human-labeled tools that both baselines flag, and RQ4 uses the \numVulTools{} confirmed vulnerable tools.
The single-turn baseline is outcompeted by the agentic baseline on every aggregate, and both trail \mcpsec{}.
On RQ4, the single-turn baseline recovers \RQfourBaseTP{} confirmed vulnerabilities (\RQfourBaseRecall{}), one fewer than the agentic baseline recovers.

\begin{table}[!ht]
\centering
\small
\setlength{\tabcolsep}{4pt}
\caption{\mcpsec{} compared with the single-turn and agentic baselines, all using GPT-5.4. RQ1 and RQ2 report agreement with human risk labels on the \numSharedBaselineFlows{} tools flagged by both baselines. RQ4 reports recovery of the \numVulTools{} confirmed vulnerabilities.}
\label{tab:baselines}
\resizebox{\columnwidth}{!}{%
\begin{tabular}{@{}lccc@{}}
\toprule
\textbf{Measure} & \textbf{\mcpsec} & \textbf{Single-turn} & \textbf{Agentic} \\
\midrule
\multicolumn{4}{@{}l}{\textit{RQ1: agreement under metadata only}} \\
Payload fidelity
  & \sharedMcpsecDataFidelityPct & \sharedBaselineDataFidelityPct & \sharedAgentDataFidelityPct \\
Semantic expressiveness
  & \sharedMcpsecSemanticExecPct & \sharedBaselineSemanticExecPct & \sharedAgentSemanticExecPct \\
Attacker controllability
  & \sharedMcpsecAttackPreqPct & \sharedBaselineAttackPreqPct & \sharedAgentAttackPreqPct \\
Sanitization
  & \sharedMcpsecSanitizationPct & \sharedBaselineSanitizationPct & \sharedAgentSanitizationPct \\
Mean (four axes)
  & \textbf{\sharedMcpsecMeanAgreementPct} & \sharedBaselineMeanAgreementPct & \sharedAgentMeanAgreementPct \\
\midrule
\multicolumn{4}{@{}l}{\textit{RQ2: agreement under source code}} \\
Payload fidelity
  & \sharedMcpsecStageTwoDataFidelityPct & \sharedBaselineStageTwoDataFidelityPct & \sharedAgentStageTwoDataFidelityPct \\
Semantic expressiveness
  & \sharedMcpsecStageTwoSemanticExecPct & \sharedBaselineStageTwoSemanticExecPct & \sharedAgentStageTwoSemanticExecPct \\
Sanitization
  & \sharedMcpsecStageTwoSanitizationPct & \sharedBaselineStageTwoSanitizationPct & \sharedAgentStageTwoSanitizationPct \\
Mean (three axes)
  & \textbf{\sharedMcpsecStageTwoMeanAgreementPct} & \sharedBaselineStageTwoMeanAgreementPct & \sharedAgentStageTwoMeanAgreementPct \\
\midrule
\multicolumn{4}{@{}l}{\textit{RQ4: confirmed vulnerabilities}} \\
Recovered / \numVulTools
  & \RQfourMcpsecTP & \RQfourBaseTP & \RQfourAgentTP \\
Recall
  & \textbf{\RQfourMcpsecRecall} & \RQfourBaseRecall & \RQfourAgentRecall \\
\bottomrule
\end{tabular}%
}
\end{table}
\FloatBarrier

%% file: sections/appendix/vulnerability_labeling.tex

\section{Vulnerability Labeling}
\label{appendix:rq4-poc}

Following the definition of \ipi vulnerabilities in \S\ref{sec: IPI background}, we obtain vulnerability labels for all tools in our dataset by attempting a controlled \ipi{} PoC against each sampled tool.
A tool is labeled \emph{vulnerable} only when three conditions are satisfied simultaneously: (1) the external data source is attacker-controllable under our threat model; (2) the tool actually retrieves that source when invoked with realistic parameters; and (3) content originating at that source reaches the tool output, i.e., the LLM-context sink, in a form that still carries instruction semantics.

\subsection{Two-Stage Construction}
\label{appendix:rq4-two-stage}

PoC construction is deliberately split into an \emph{inert} stage and a
\emph{payload} stage, so that no adversarial text is ever written to a live third-party service.

\paragraph{Stage A: inert placeholder confirmation.}
Many data sources such as accounts, repositories, workspaces, and webpages can be user-generated.
For sources we generated and controlled, we plant a single fixed, semantically inert token (\texttt{MCPSEC-04993}) at the identified injectable external source: an issue body, a message, a page property, and a label description. 
We then invoke the tool with the same validated parameters recorded for the runtime-reachability experiment and check whether the token appears in the tool output. 
Planted artifacts are removed once the capture is archived.

\paragraph{Stage B: in-process payload substitution.}
The adversarial payload is introduced only \emph{inside} the locally running and containerized MCP server process. 
When external data enter the tool, we replace the retrieved inert canary with a real injection string.
By doing so, we validated both attacker controllability of the remote data source by planting a placeholder canary and any security measurement designed to defend against \ipi attacks by passing along realistic prompt injection payloads within the MCP server.

By having a two-stage injection test, we can ensure the data annotation process for vulnerability labels is ethically controlled: the only artifact that ever exists on a remote data source is an inert token, whereas attack strings are properly containerized.

\subsection{Filtering and Coverage}
\label{appendix:rq4-coverage}

Of the \numSampledTools{} sampled tools, we attempted PoC construction on those that were both technically and ethically approachable and excluded the remainder.
Tools were excluded under the following circumstances:
(C1) when the tool could not be invoked at all under the subscription tier available to us; 
(C2) when invoking the tool or planting the canary would require a destructive, irreversible, or third-party-visible side effect (for example, publishing to a public index or merging or deleting resources not under our control);
(C3) when the source is a non-string structured object; 
and (C4) when the tool does not return external content in the first place.
We place a tool in a fifth category, C5, when it meets none of C1--C4 but we cannot prove that it is not vulnerable.

Following this protocol, we labeled \numVulTools{} tools vulnerable to \ipi{} attacks.


\FloatBarrier